# Functional Decision Theory in an Evolutionary Environment

**Noah Topper, University of Central Florida**

## Abstract

Functional decision theory (FDT) is a fairly new mode of decision theory and a normative viewpoint on how an agent should maximize expected utility. The current standard in decision theory and computer science is causal decision theory (CDT), largely seen as superior to the main alternative evidential decision theory (EDT). These theories prescribe three distinct methods for maximizing utility. We explore how FDT differs from CDT and EDT, and what implications it has on the behavior of FDT agents and humans. It has been shown in previous research how FDT can outperform CDT and EDT. We additionally show FDT performing well on more classical game theory problems and argue for its extension to human problems to show that its potential for superiority is robust. We also make FDT more concrete by displaying it in an evolutionary environment, competing directly against other theories. All relevant code can be found here: https://github.com/noahtopper/FDT-in-an-Evolutionary-Environment.

# 1 - Introduction

Here we will lay down the basics of what it means to maximize expected utility, why it is that different theories can disagree about how to do it, and what exactly makes FDT special and useful. Much of this work has been done in the paper introducing FDT: *Functional Decision Theory: A New Theory of Instrumental Rationality* (Soares and Yudkowsky 2018). But the explanations are included here for the sake of completeness, and also to explain the concepts in simple terms that will hopefully be readily understandable to the naive reader or the more classically-minded decision theorist. The reader who is already familiar with FDT may gain some new insights in this section, but may instead wish to simply skip on to the next. In future sections, we will set up and use an evolutionary environment for FDT to compete against other decision theories.

## 1.1 Basic Utility Theory

In 1947, it was proven (Neumann and Morgenstern 1953) that any agent following a simple set of axioms has a utility function that describes their preferences and behaves so as to maximize the expected value of this function.

These preferences are in regard to all possible outcomes. An agent might prefer an outcome in which they have more money, for example. Any particular event we are concerned about can be an outcome evaluated by the utility function.

The utility function is a function in the fairly straightforward mathematical sense: each outcome maps to a single utility value. This represents how much the given agent values that outcome. Other possible outcomes may have lower, higher, or equal utility.

In addition, this is a cardinal utility function. This means that the higher a utility value is, the *stronger* the preference, and that we can infer relatives strengths of preferences from these numbers. For example, "100 utility" has a consistent meaning. The difference between 0 utility and 100 utility has the same weight as the difference between 100 and 200. On this scale, a rational agent acts so as to maximize their own expected utility.

There are four axioms that an agent must follow to have such a function, which are defined below. Here, $A$, $B$, and $C$ are different possible outcomes, and the symbol $\geq$ represents a preference order, so $A \geq B$ means that outcome $A$ is at least as good as $B$, if not better, according to a particular agent's preferences. It is also important to note that combinations of outcomes, like $pA + (1 - p)B$, represent mutually exclusive lotteries. That is, a probability $p$ of getting outcome $A$ and otherwise getting $B$: always one and never both.

- Completeness: For all possible outcomes $A$ and $B$, either $A \geq B$ or $B \geq A$. That is, either $A$ is preferred to $B$, $B$ is preferred to $A$, or the agent is indifferent between the two, in which case both expressions are true.
- Transitivity: If $A \geq B$, and $B \geq C$, then $A \geq C$. That is, if $A$ is better than $B$, and $B$ is better than $C$, then $A$ must be better than $C$.

- Independence: If $A \succcurlyeq B$, then for any probability $p$, $pA + (1-p)C \succcurlyeq pB + (1-p)C$. That is, if $A$ is preferred to $B$, then a mixing of $A$ and $C$ should be preferred to the same mixing of $B$ and $C$, regardless of what $C$ is.

- Continuity: If $A \succcurlyeq B \succcurlyeq C$, then there exists a probability $p$ such that $pA + (1-p)C$ is equally as good as $B$. That is, if $A$ is better than $B$, and $C$ is worse than $B$, some mixing of $A$ and $C$ must balance out to be just as good as $B$.

These axioms may seem sensible, but they also might not seem entirely beyond doubt. Morgenstern and von Neumann showed that any agent which does *not* follow these axioms would be willing to agree to a Dutch book, which is a series of bets that is a guaranteed loss. So if by "rational", we mean an agent being able to achieve its preferences consistently, then no agent that does not follow these axioms can be rational. This will be a basic assumption moving forward.

So what does it mean to maximize expected utility? First, what is expected utility? If an agent takes a bet that has a 0.5 probability of resulting in 100 utility and a 0.5 probability of 0 utility, then the expected utility is $0.5(100) + 0.5(0) = 50$ utility.

In general, we have:

$$EU(a \mid x) = \sum_{i=1}^{N} P(a \to o_i \mid x) \cdot U(o_i),$$

which says that the expected utility of an action $a$ is the sum over all the possible outcomes of that action, weighted by their probabilities and utilities. $P(a \to o_i \mid x)$ is the probability that taking action $a$ will lead to the particular outcome $o_i$, given some background information and/or evidence *x*, which could simply be the problem description. The *x* should include all relevant factors, including all information about other players, so this covers decisions and games (problems involving other agents). *U(o_i)* is then the utility of that outcome.

Maximizing expected utility given background information *x* then means selecting an action[1] *a\** that leads to the highest result of this equation. The issue of how this relates to pure versus mixed strategies (i.e. randomly selecting between certain actions) will be handled in the next section. So we want the output of our decision process to be:

$$a^* = \mathop{argmax}\limits_{a} \sum_{i=1}^{N} P(a \to o_i \mid x) \cdot U(o_i),$$

where *argmax* means selecting the *argument* that *maximizes* the function we have defined.

Our overall parameters are a set of actions *A*, a set of outcomes *O*, some data/background information *x*, and a utility function *U* over the outcomes. Our tricky parameter is *P*, corresponding to the $P(a \to o_i \mid x)$ term. We need a probability distribution over how actions lead to outcomes. The different theories we wish to consider in this paper actually disagree about how such a *P* should be defined or constructed, so we cannot simply take it as given. We must directly confront the issue of how to build such a *P* from the information we have, *x*. This issue is rooted in the problem of counterfactual reasoning.

---

[1] Different actions may be equally good. Assume we have some arbitrary rule for settling ties.

## 1.2 Counterfactual Reasoning

Counterfactual reasoning is the act of imagining alternatives, considering what *could* have happened, rather than what *did* happen. Often, it particularly refers to the thought of "What would have happened if I had acted differently?" Extending this notion slightly, when making a decision *right now*, one must consider what *would* happen given each available choice, until settling on one choice that determines what actually *will* happen.

These other actions considered, then, are essentially the same as counterfactuals: one must determine the consequences of actions that are never actually taken. *P* is nothing but a description of what would happen given different actions, so it is built from these counterfactuals. In this section we will also try to make it clear that this reasoning is more than just analogous to counterfactual reasoning, it really is the same thing.

Counterfactual reasoning can clash with the idea of determinism, the notion that everything that occurs is fully determined by past events, making even the future itself already determined by the present. If our world is deterministic, then asking how things could have happened differently is somewhat confused, since things could *not* have happened differently. If we rewound a deterministic world and let it play forward again, everything would happen just the same. This is not to say that one's actions do not have consequences, just that actions themselves may be determined. This is relevant because, regardless of how the world may be, a good decision process *should* be deterministic.

To see why, imagine that we have found a way to actually specify and implement an algorithm that maximizes expected utility, as described above. We will have designed a decision algorithm (or process, function, etc.): a process which, when faithfully executed, takes in a description of a problem along with a set of observations, which together we may call $x$, and outputs a decision or action, which we may call $a^*$. This means iterating over all potential actions $a$, and choosing the action that maximizes $EU(a|x)$.

The important thing to note here is that if such a concrete process were given an identical problem at a different point in space or time, it would output the same result. This is the essence of determinism: if we give the decision algorithm the same input, it will always produce the same output.[2]

Two points of clarification are in order. First, one may object that many decision algorithms occasionally call for randomness. In many game theory problems, randomly selecting between actions, or mixing, is the optimal strategy. If two agents using precisely the same decision algorithm played a game against each other, and both decided to determine their actions based on the result of a coin flip, they could execute different actions even given the same game. Nonetheless, the choice to flip a coin was arrived at deterministically, and so this should always be the output of the decision algorithm. Our action set $A$ therefore includes both pure and mixed actions.

Second, we also get randomness from Nature. When an agent takes an action, they are not entirely certain what the outcome will be, due to the unpredictably of Nature. But it is the outcomes here that are random, not the agent's choices. *P* accounts for both of these sources of randomness. Overall, when our agent takes an action, it may be a mixed action that probabilistically leads to pure actions, while pure actions themselves probabilistically lead to certain outcomes. But none of this violates the assumption of a deterministic decision process. See Figure 1.2.1 for an example of what *P* is doing.

---

[2] We are assuming here that space and time are not directly relevant in the problem at hand, so one is not rewarded by fiat for executing the decision algorithm at a certain place or time. So $x$ only contains problem-relevant data. It is also possible that the environment $x$ is dependent on space and time. This is fine, we just assume that once we fully know $x$, space and time are irrelevant. If we have all the same problem-relevant characteristics at a different place and time, we should get the same output from our decision process.

**Figure 1.2.1**

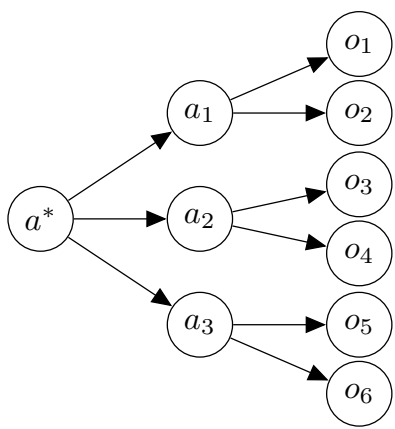

Here, our optimal action $a^*$ happens to be a mixed action, randomly leading to one of three pure actions, $a_1$, $a_2$, or $a_3$. Individually, each of these randomly leads to one of two outcomes. *P* is taking in $a^*$ and performing these operations to get the probabilities of each outcome in the last column.

So even with mixing, an optimal decision algorithm should be deterministic. Earlier we claimed this clashes with counterfactual reasoning. Determinism tells us that, with the same input, our decision algorithm should always produce the same output. Counterfactual reasoning, however, requires that we consider taking alternative actions. To see the arising clash clearly, let us say that our agent is in a given problem described by *x*. It happens to be the case that action *b* from *A* is the only optimal choice, and the agent will therefore choose that action (given that its decision function really is optimal). But of course the agent does not know that in advance, while it is still considering its choices.

So the agent considers, "What would happen if I choose *c*?" What makes this a strange and difficult question is that we in fact know (on the outside) that *b* is the only possible output from the decision function on the given problem, since the function is optimal and deterministic. If we call this function *f*, the agent is asking, "What would happen if f(x) = c?" when in fact f(x) = b. It is not actually possible that f(x) = c, and yet to analyze our options, we must consider it anyway. If we want to know the expected utility of *b* and *c*, the method we have defined necessitates first calculating the expected *outcomes* of taking actions *b* and *c*. In particular, we must determine the implications of f(x) = c.

This also explains why counterfactual reasoning before and after a decision are not truly different. Afterward, the agent is asking, "What would have happened if, when I was in problem *x*, I had chosen *c*?" Beforehand, the agent is asking, "What will happen if, in this problem *x*, I choose *c*?" Both of these are mathematically equivalent to asking "What if f(x) = c?", and so there is no distinction between the two.

This is counterfactual reasoning, because we are reasoning about things that are *counter to the facts*. When one considers "What if f(x) = c?" this is counter to the fact that f(x) = b. Before the choice is made, it may seem *conceptually* possible to the agent that f(x) = c, when in fact it is physically and logically impossible. Counterfactual reasoning is therefore a *conceptual* tool. It may be impossible that f(x) = c, but we hope it is possible to derive its implications anyway. In fact, what we infer from all this is that *if* our decision function is optimal, then it *must* be dealing with these questions somehow.

One may say that what will happen if we take action *c* is defined by *P*, since it contains the consequences of taking every available action. This is true, but our task now is to determine how to construct *P*. To take *P* as given would be to ignore the problem at hand. *P* describes the outcomes of taking various actions, so this is where all the counterfactual reasoning occurs in our formula. To build it, we need a reasonable way to move from our data *x* to a *P* that reflects sensible counterfactuals.

Some may wish to brush aside talk of counterfactuals as overly philosophical, but it is hard to see how to construct a satisfactory *P* without considering the problem. In fact, all that *P* does is answer the question "What would happen if I take action *a*?", alternatively written as "What would happen if f(x) = a?", when in fact f(x) may equal something else. *P* is therefore *nothing but* a collection of counterfactual statements. Constructing an adequate *P* then must be how our optimal algorithm is doing what it does.

To summarize, we suppose *b* is optimal. Then if *f* is also optimal, f(x) = b. This optimal algorithm must calculate the expected utility of other actions, such as f(x) = c. This means the optimal algorithm must be considering counterfactuals. Finally, since *P* contains all our counterfactual statements, our optimal algorithm is dependent on constructing an adequate *P*.

What would it look like to solve this problem? If our agent imagines a world where it took action *c*, this counterfactual world is obviously different from the real, factual world. Everything in the real world led to action *b*, therefore if action *c* had been chosen, the world must have had different qualities to lead to this different outcome.

What differences, exactly? That is the difficult question with different perspectives. Each of the three theories we consider essentially constructs a hypothetical (or counterfactual) world model *x'*. The first thing each theory does is take its current world model *x* and tack on the given action it is considering taking, *a*. Each theory then makes further modifications to this hypothetical world to accommodate the given action. Finally, each theory calculates in a straightforward manner what would occur in this hypothetical world.

This is actually what $P$ is doing. To determine the consequences of action $a$ given background information $x$, $P$ builds a hypothetical world by combing $x$ and $a$ in some manner, and calculating the consequences. Each theory, then, has different methods for constructing $P$, which really means different methods for building hypothetical worlds.

As a small preview, evidential decision theory makes large modifications to $x$, causal decision theory makes minimal modifications, and functional decision theory is somewhere in between. Bear in mind that as we describe each such process, we are also essentially defining $P$ under that theory by laying out the operations that $P$ performs.

The only way to make effective choices is to consider one's options. To consider its options, each theory will imagine hypothetical worlds, different from the real world. These worlds will be *conceptually* possible, but perhaps not *literally* possible, and so they may clash with our intuitions. We could have philosophical squabbles over which of these theories makes the most metaphysical sense, or we could simply go for pragmatism. All of these theories are attempts at maximizing expected utility, so whichever does this the most effectively ought to be the theory we go with, unless another theory later surpasses them all. So let us now concretely examine how these theories differ.

# 1.3 Introduction to EDT, CDT, and FDT

All three of these theories prescribe different ways to deal with counterfactual reasoning in an attempt to maximize expected utility. They can also be seen a prescription for constructing a decision algorithm, a set of step-by-step instructions that can be followed and executed to make decisions and solve problems. We could imagine it as an elaborate list of instructions that a human could follow to the letter, or we could theoretically design an agent, such as a computer program or artificial intelligence (AI), that has these instructions inbuilt.

**Evidential Decision Theory**

Evidential decision theory (EDT) is the simplest theory here. To determine the consequences of an action, it simply asks, "What would happen *given* that I take action *a*?" This means it treats $P(a \rightarrow o_i \mid x)$ as $P(o_i \mid x, a)$, which translates to "the probability that taking action *a* will lead to outcome $o_i$ is simply the probability that outcome $o_i$ occurs, given that I took action *a*." This probably seems quite intuitive. The equation is then:

$$EDT(x) = \underset{a}{argmax} \sum_{i=1}^{N} P(o_i \mid x, a) \cdot U(o_i).$$

EDT is treating counterfactuals as pure conditionalization, meaning it conditions on each possible action, taking it as given, and outputs the action that corresponds to the highest expected utility. This conditioning is where the term "evidential" comes from. Normally we put our pieces of evidence after the | symbol. EDT is treating its own actions as pieces of evidence to update on. Thus it is also sometimes summarized by the question, "What choice would be the best news to hear I took?"

When EDT conditions on a particular action *a*, it then updates its beliefs about everything that is statistically associated with *a*. It had a given world model *x*, but of course with any new piece of information, *x* should be updated. EDT updates all values contained in *x* that are correlated with its action. This is also a description of how EDT constructs its hypothetical world. It forms *x'* by first including *a* in *x*, and then updating all associated values. This is its probabilisitic model.

There are problems with updating all these values. Take the following example scenario. Tobacco companies previously claimed that, in spite of the statistical link between smoking and cancer, we could not be certain that the link was causal rather than merely a correlation (Clive and Rowell 2004). For example, what if the reason cigarette smoking and cancer are correlated is because there exists a gene which makes smoking more addictive and also makes one more susceptible to cancer?

Imagine that we live in the world where this is true and known to all. What ought one to do? If one enjoys smoking and it does not actually *cause* any negative side effects, it seems one would benefit from doing so. But from the point of view of EDT, *given* that the agent smokes, they are much more *likely* to develop cancer. Therefore when EDT imagines smoking, it updates its probability of having the smoking gene and developing cancer. EDT then avoids smoking simply because it is correlated with bad outcomes. One might say EDT confuses correlation with causation.

This reasoning may even seem appealing to some readers. Bear in mind that this reasoning would just as easily tell one to avoid coffee just because some people are born with a gene that makes them more likely to enjoy coffee and to develop cancer. One either has the gene or they do not. Either way, they only lose by avoiding the coffee or cigarettes now.

When an EDT agent imagines taking a particular action, it imagines the whole world shifting to accommodate this action, potentially even the agent's own gene profile. This is clearly physical nonsense, but more importantly, it does not work practically. It generally does not perform very well. In contrast, CDT and FDT both maintain that most of the world remains stable when they imagine making different choices.

EDT is generally not very popular in the field of decision theory. Why would anyone want to use it? Mainly because it is simple. Determining causation is a difficult task, while identifying correlations is rather easy. This is really just a pragmatic concern of actually building a system that maximizes expected utility. Perhaps it would be too difficult to build something that uses causal reasoning, or if we want to solve a simple problem, EDT might work well enough.

For example, imagine we have a vacuum robot, operating on a grid, that has five possible actions: Up, Down, Left, Right, and Suck. We program the robot to believe that executing Suck sucks up the dirt in its current location with probability 1, Right leads to moving right with probability 1, etc. We make this robot's goal to suck up all the dirt on the grid in as few moves as possible.

If we give it no other knowledge, EDT works just fine. Given that it executes Right, it updates its belief about its location, and nothing else. This is because its actions are only associ-ated with the relevant outcomes. We have no need for a complex causal model on such a simple problem. It is when we wish to develop complex reasoning and consider counterfactuals that it falls apart.

Additionally, that are some more complex problems, such as Newcomb's Problem which we will discuss later, that EDT seems to perform *better* on than a theory that incorporates causal reasoning. This is a reason that evidential decision theory and causal decision theory are sometimes seen as competing theories.

**Causal Decision Theory**

Causal decision theory (CDT) is a significant change from EDT. Its policy for dealing with counterfactuals is essentially a total inversion. It says that since the only thing the agent can control is its own actions, when imagining taking a particular action, they should imagine the entire world staying exactly fixed, with only the action varying, and of course the direct physical consequences of the new action. CDT therefore asks "What physical act of mine will cause the best outcome, assuming the rest of the world stays fixed?" There a few ways to write this out mathematically, but one is:

$$CDT(x) = \underset{a}{argmax} \sum_{i=1}^{N} P(o_i \mid x, \mathbf{do}(a)) \cdot U(o_i),$$

which says we output the action that maximizes expected utility given that we **do**(*a*).

What is this **do**( · ) operator? It comes from computer scientist Judea Pearl's work in the formalization of causal reasoning, for which he won the Turing Award. See *Causality: Models, Reasoning, and Inference* for more information.

The **do**( · ) operator means we first set our action to a particular value, just like before with EDT. But additionally, we "disconnect" our action from all its prior causes. So it is a two-step process. This means that if we take **do**(*a*) as given, we only update a *subset* of the values in our world model *x* that are correlated with *a*. In particular, we do not update any of the *causes* of *a*, we only update the *effects* of *a*. This is how CDT constructs its hypothetical worlds.

But how do we determine which of our variables are causes or effects of *a*? Pearl's work includes methods for isolating causal relationships from statistical data. Starting with our knowledge *x*, we use these methods to construct a causal model of the data, which can then be utilized by the **do**( · ) operator. This requires an *x* that is fairly rich in data. So we build *P* from *x*.

Causal models have a very elegant graphical representation, again from Pearl. Write every variable in the model as a node. Draw a directed edge between two nodes if the first exerts a causal influence on the second. These edges are weighted by the strength of the correlation. To **do**(*a*) means finding the node representing the value we wish to change, set it to *a*, remove its incoming edges, and update all connected values. Consider the following graphs.

**Figure 1.3.1**

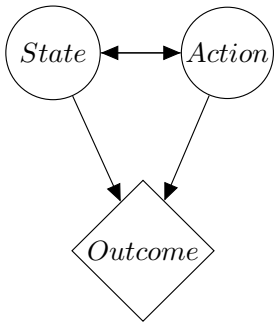

Let this graph represent *x* in general. **State** holds our knowledge of the current environment, and the arrows represent our knowledge of the correlations between our variables. The arrows between **State** and **Action** are bidirectional, since we know they are correlated, but *x* does not immediately tell us which direction the causal influence goes. Let us assume, however, that we can treat **Outcome** strictly as a result of the state and our action.

**Figure 1.3.2**

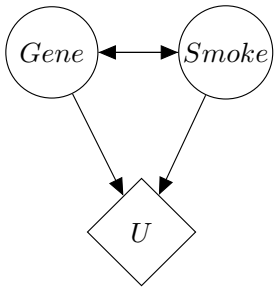

Here is a possible *x* for the smoking problem and our given agent. The state **Gene** is the probability that the agent has the smoking gene. The action **Smoke** is the probability they will pick up smoking. The outcome is their **Utility**. From population data, 75% of smokers have the gene, 10% of non-smokers have the gene, 80% of those with the gene develop cancer, and 20% of those without the gene develop cancer. Cancer results in -100 utility for the agent, and smoking results in +5 utility. We could have **Cancer** as an intermediary node between **Gene** and **Utility**, but we capture the relation in the statistical connections for simplicity of the diagram.

**Figure 1.3.3**

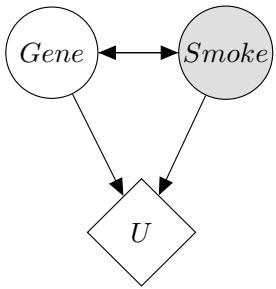

Our agent is deciding for the first time whether or not to start smoking. We consider smoking and not smoking, and compare the expected utilities. Under EDT, we first take it as given that we smoke, setting **Smoke** = 1. We update the value of each variable connected to **Smoke**. **Utility** increases by 5, and our probability of having the gene becomes 0.75. The probability of getting cancer due to having the gene is 0.75 x 0.8 = 0.6. The probability of getting cancer due to not having the gene is 0.25 x 0.2 = 0.05. Together there is a 0.65 probability of getting cancer, resulting in -65 expected utility, for a net expected utility of -60.

We then consider not smoking, **Smoke** = 0. Our probability of having the gene becomes 0.1. The probability of getting cancer due to having the gene is 0.1 x 0.8 = 0.08. The probability of getting cancer due to not having the gene is 0.9 x 0.2 = 0.18. Together there is a 0.26 probability of getting cancer, resulting in -26 expected utility.

According to EDT, not smoking is the clear choice. But EDT has made an error, believing that smoking somehow retroactively changes its own genes. They are correlated, but **Smoke** exerts no causal influence on **Gene**.

**Figure 1.3.4**

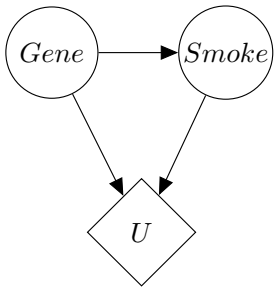

For CDT, we have this modified version of *x*, a causal model. The directed arrows represent directions of causal influence. The agent's genes causally influence their chances of developing cancer and their chances of choosing to smoke. Smoking, however, has no causal influence on one's genes. We otherwise assume the same relationships among the variables and population data as before.

**Figure 1.3.5**

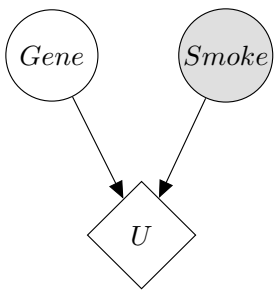

Under CDT, we first consider smoking, taking **do(Smoke** = 1) as given. This operator first deletes the edge from **Gene** to **Smoke**. This prevents any alterations to **Smoke** from affecting **Gene**. Then we set **Smoke** = 1. This increases **Utility** by 5.

We then consider not smoking, taking **do(Smoke** = 0) as given. **Utility** is unaffected. In both these cases, our probability of having the gene and developing cancer remains stable. According to CDT, smoking is obviously superior.

This seems like an improvement, and indeed it is. EDT's hypotheticals are strange because they imagine significant changes in the world resulting from actions with no causal connection to the altered variables, and it turns out this leads to problems. CDT avoids these problems, but its hypotheticals are strange in their own right.

When CDT constructs a hypothetical world, it disconnects its actions from all possible inputs. This is rather strange, because one's actions *do* depend on many things. This can lead to problems of its own. The classic example in this area is Newcomb's Problem (Nozick 1969).

In Newcomb's Problem, an agent is faced with two boxes. There is a transparent box clearly containing $1,000 and an opaque box that is known to either contain $1,000,000 or $0. The agent may either take both boxes, or just the opaque box. The catch is that the boxes were

filled by a reliable predictor who has been correct on 99% of previous games. The predictor analyzed the agent in advance, perhaps with a brain scanner, and placed the $1,000,000 inside the opaque box if and only if it predicted that the agent would only take the opaque box. What should our agent do? Take both (two-box) or just the opaque box (one-box)?

Causally speaking, the prediction has already been made, and the opaque box is now sitting there, either full or empty. Nothing the agent does *now* can change the current state of the box. Whether it is full or empty, two-boxing is obviously superior.

Yet, if our agent reliably reasons in this manner, the predictor will likely be able to determine this in its analysis. So our agent will typically end up with only $1,000. If our agent settled on one-boxing, the predictor should know this as well, and our agent would typically end up with $1,000,000.

**Figure 1.3.6**

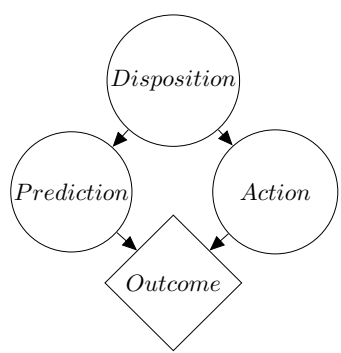

Here is a causal graph of the setup we have described. The agent's **Disposition** influences which **Action** they will take (one-box or two-box). This simply means that the way the agent reasons determines their action. Their **Disposition** also influences what **Prediction** will be made by the predictor. Together, these two values determine the **Outcome**, the sum of money the agent ends up with. Formally, **Disposition** could take on values from the space of all possible decision functions, and **Prediction** is some probability $p$ that the predictor anticipates the agent two-boxing. **Disposition** and **Prediction** together sum up the state of the game.

**Figure 1.3.7**

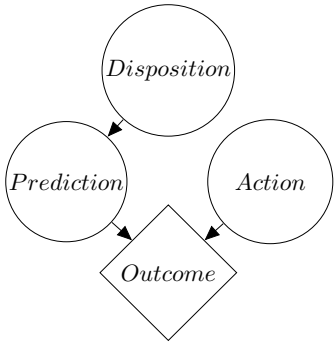

To determine the best choice, CDT severs the connection between its **Disposition** and its **Action**. Then it tries setting **Action** equal to "One-Box" then "Two-Box" to see which leads to a better result. We can see that **Prediction** remains fixed. **Prediction** is some value $p$, the estimated probability that the predictor will anticipate the agent two-boxing. The expected value of one-boxing is then $p(\$0) + (1-p)(\$1{,}000{,}000)$ and the expected value of two-boxing is $p(\$1{,}000) + (1-p)(\$1{,}001{,}000)$. Regardless of $p$, two-boxing seems superior.

Since CDT certainly two-boxes, and the predictor has a 99% chance of anticipating this, let $p = 0.99$ Then CDT's expected value on this problem is $0.99(\$1{,}000) + 0.01(\$1{,}001{,}000) = \$11{,}000$. But EDT reliably one-boxes on this problem. It reasons "Given that I one-box, I expect there to be $1,000,000 inside the opaque box with probability 0.99. Given that I two-box, I expect $0 to be in the opaque box with probability 0.99. So I will one-box." With probability 0.99, the predictor knows that EDT will one-box, so let $p = 0.01$. EDT's expected value on this problem is then $0.01(0) + 0.99(\$1{,}000{,}000) = \$990{,}000$.

CDT is clearly beat out by EDT on this problem. Why does this occur? When CDT disconnects its **Action** from its **Disposition**, it is treating its own action as though it were totally independent of all causal influence, because it allows no causal inputs into its **Action** node whatsoever. This leads to **Prediction** being held fixed. To sum up CDT's particular failure mode here, *CDT assumes its actions are independent all of predictions made about it*.

A CDT agent has some beliefs in *x* about how its opponent will model it. It holds all these guesses fixed as it imagines taking various actions.[3] It therefore behaves as if its actions are totally independent from any predictions made about it. If this predictive modeling is being done accurately, the agent will quickly be outwitted.

In general, CDT has no way to represent the fact that some decision functions are correlated with its own. The predictor's decision is correlated with CDT's decision, for example. If another decision function reliably produces outputs correlated to CDT's outputs, but neither exerts a direct causal influence on the other, CDT will always assume that the outputs of the two functions are independent. This problem can potentially extend to any game in which opponents model each other at all!

CDT seems to be such a sensible policy! Why does it go wrong? One simple answer is that CDT has no model of its own decision function in its causal map of the problem. How can it tell that some actions are the outputs of correlated decision functions if it never bothers representing decision functions in its model at all? All actions are represented as sourceless, so a decision function has no place in its model. It has no way to tell that it is making these errors, because it cannot represent the fact that different actions are dependent on the same decision function.

---

[3] In game theory, this is the assumption when searching for a Nash equilibrium: hold the opponent's action fixed as one considers their own choices. This is why CDT is the formal framework of most of modern decision theory.

**Functional Decision Theory**

It seems that EDT changes its world model too drastically when considering hypotheticals, while CDT holds it too fixed. Functional decision theory (FDT) is an attempt to find an optimal middle ground. Could we possibly find some set of variables in our world model $x$ which we update in $x'$ that leads to truly optimal behavior? FDT claims it can. Let us see how.

First, FDT is an attempt to incorporate the idea of embedded agency into decision theory. Embedded agency means that an agent is not separate from the environment it is acting in, but is instead part of the environment itself. Our agent wants to choose the best possible action, but whatever action it chooses is just another fact about the environment. In a CDT agent's causal model, the agent itself is "outside" the environment. All of its inner-workings and decision processes are not represented in the model, as if the agent and the environment were two entirely separate things. We call agents like this dualistic. This is *not* how an FDT agent models its environment. In FDT's model, the agent's own decision process is placed directly in the causal model, and the agent reasons accordingly.

To do this, FDT starts with its information $x$ and builds a causal model out of it, just like CDT does. Additionally, its model must represent the fact that some variables are dependent on the decision procedures of agents, and that some decision procedures are correlated. So we again need a richer $x$ than before. The biggest open problem in FDT is that we currently have no method for detecting such dependencies from raw data. We can still deal with simple problems where the EDT, CDT, and FDT models are straightforward, but progress in this area could be quite useful.

The equation for FDT is

$$FDT(x) = \underset{a}{argmax} \sum_{i=1}^{N} P(o_i \mid x, \mathbf{do}(FDT(x) = a)) \cdot U(o_i),$$

which says we output the action that maximizes expected utility given that we **do**$(FDT(x) = a)$.

This has a similar meaning to before, but now we have a variable in our model that represents our agent's own decision function. We call it **FDT(x)**, and it represents the outcome of our decision function on the given problem. We are treating it as a variable because we do not yet know what its value will be.

While before we might have set Smoke equal to *a*, now we set **FDT(x)** equal to *a*, and calculate all the downstream consequences, similar to before. That is, in our new world model *x'*, we set **FDT(x)** to a hypothetical value, and update the values of all the variables that are logically dependent on the outcome of our decision function.

What do we mean by "logically dependent"? This term essentially stands in for what we said before, that some variables are dependent on agent's decision procedures, and that some decision procedures are correlated. So we update our beliefs about the outputs of decision functions correlated with our own, and we update our beliefs about variables that these decision functions control. Graphically, we can see what this modified process looks like below.

**Figure 1.3.8**

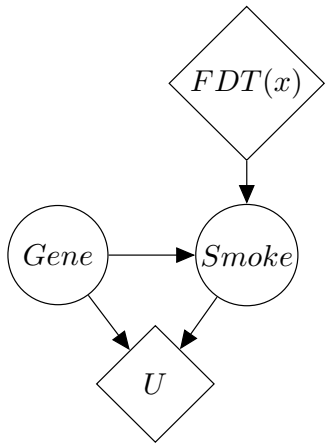

We have a causal model similar to before. We augment it by noting that the agent's choice of whether or not to smoke is influenced by its decision function. The decision function itself is taking the whole structure of the problem *x* as its input. The other relationships remain the same as before.

**Figure 1.3.9**

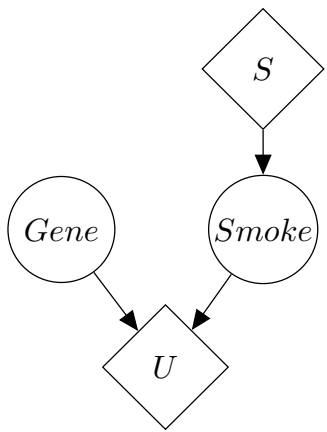

When considering an action, FDT intervenes at **FDT(x)**, rather than **Smoke**. If S stands for the act of smoking, we consider it by taking **do**(FDT(x) = S) as given. Under FDT, when we do this, we remove all incoming edges from our action node except for **FDT(x)**. That is, we treat the decision function as the sole cause of the action.

FDT iterates over each possible output of the decision function (here S and ¬S). All descendants of **FDT(x)** are then updated. Here, **Smoke** is the only connected variable. So **do**(FDT(x) = S) simply sets **Smoke** = 1 and **do**(FDT(x) = ¬S) sets **Smoke** = 0. The result, then, is exactly the same here as under CDT.

This may not seem any better than CDT, since the agent's action is still severed from the state of the problem. This is not quite true, however. The agent's action is dependent on its decision function, which itself takes the whole state of the problem as its input.

Why can this not happen under the other frameworks? Under EDT, actions are not special variables: they are dependent on simple correlations between the other state variables. Under CDT, actions are quite special, in that they are treated as not dependent on any other state variables: they are the result of intervention from outside the model. Under FDT, actions are also special, in that they are the only variables that are dependent on decision functions. FDT is the only framework that includes such dependencies, and it is therefore the only one that can model actions being dependent on the whole structure of the problem.

Also, in the example above, since **Smoke** is the only variable that is dependent on **FDT(x)**, intervening at **FDT(x)** produces the same result as simply intervening at **Smoke**. This is true in general: if only one variable depends on the agent's decision function, the agent's action, then FDT behaves just like CDT. The key difference is that if *multiple* variables are dependent on the outcome of the agent's decision function, FDT will update the values of all these variables in its hypothetical world model. For an example, let us return to Newcomb's Problem.

**Figure 1.3.10**

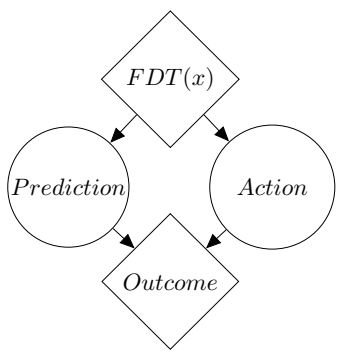

Under the lens of FDT, we replace **Disposition** with **FDT(x)**. While before we intervened at **Action**, we now intervene at **FDT(x)**. The agent's **Action** is dependent on the output of its decision function, but so is the predictor's **Prediction**, because the predictor analyzes how the agent makes decisions to come up with its prediction.

Setting **FDT(x)** = "Two-Box" sets **Action** = "Two-Box" with probability 1, but it also sets **Prediction** = 0.99. That is, there will be a 0.99 chance that the predictor anticipates the agent two-boxing. This results in an expected value of $11,000, as calculated before.

Setting **FDT(x)** = "One-Box" sets **Action** = "One-Box" with probability 1, and sets **Prediction** = 0.01. This results in an expected value of $990,000. So FDT one-boxes.

Since FDT models the fact that actions depend on decision functions, it can also model when multiple variables are dependent on the *same* decision function. In this case, it realizes that its action and the predictor's prediction both depend on its decision function, and it achieves better results from this understanding. We will see further examples in the next section.

FDT therefore succeeds on both of our example problems, unlike EDT and CDT. While one might have been tempted to categorize EDT's success as an edge case where irrational behavior leads to a beneficial outcome, FDT is a principled way to achieve success in both problems and seems to work well in general. CDT's failure relative to FDT therefore cannot be as easily dismissed.

Like the other theories, FDT's hypotheticals are rather strange. Similar to EDT, it is possible for the values of some already observed variables to change drastically in FDT's hypothetical world model. FDT still claims it holds the optimal middle ground because it does this far less than EDT does, and because it only makes changes that seem logically necessary by the constraints of determinism: deterministic functions must have the same outputs on the same problems.

We could argue that this is the most sensible way to imagine counterfactuals. Its counterfactual reasoning will begin to look rather unintuitive on some problems, however, so the real question is, does it work? The answer appears to be yes, so let us go over the advantages it gives us.

## 1.4 Qualities of FDT

### "Twin" Dilemmas

As previously noted, FDT's alternate perspective is only relevant compared to CDT when multiple variables in the given problem are dependent upon the same decision function. This is most clearly seen when two agents in a problem use precisely the same decision function.

These agents are then "twins" or "copies" of each other. We also imagine that this fact is common knowledge between the participants. Adding this assumption to the most basic of games can produce interesting results. Take the classic Prisoner's Dilemma as an example.

**Figure 1.4.1**

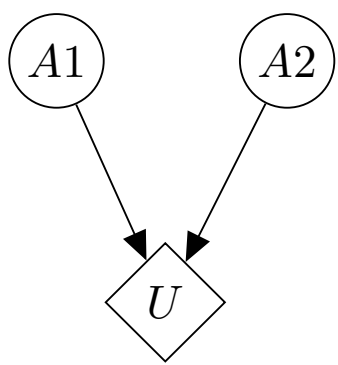

|   | Agent 2 | |
|---|---|---|
|   | C | D |
| C (Agent 1) | 7, 7 | 1, 10 |
| D (Agent 1) | 10, 1 | 4, 4 |

Here is an example of a Prisoner's Dilemma in the classical, CDT framework. The payoff matrix shows two agents facing off against each other. If both agents choose to cooperate with the other (C), they receive 7 utility. If both choose to defect against the other (D), they receive 4 utility. If one defects while the other cooperates, the defector receives 10 utility and the cooperator receives 1.

The CDT process prescribes iterating over each potential action while holding the opponent's action fixed, and selecting the one that results in the highest-utility outcome. We can capture this with a causal digram as well. The utility received depends upon the actions of both agents. The given agent intervenes at its own action node **A1** while leaving **A2** fixed. Doing so here reveals that, regardless of the opponent's action, defecting is the better choice.

CDT always prescribes defecting in this scenario. The problem with that is, if both agents are using the same decision algorithm being implemented on the same problem, the agents must output the same results. If the first agent's reasoning led it to cooperate, its twin's reasoning would necessarily do the same, and both agents would receive 7 utility instead of 4. When the agent imagines itself cooperating while its opponent still defects, it is not imagining a genuine possibility.

CDT's graph does not model the fact that both agents' actions depend on the same decision function. Even if it did, the given agent would sever and ignore any such connection in the process of making a choice, as it does with all such causal inputs. This means that even if a CDT agent were truly in a scenario competing against a copy of itself, it would be incapable of leveraging that information.

Let us reanalyze the problem under the lens of FDT.

**Figure 1.4.2**

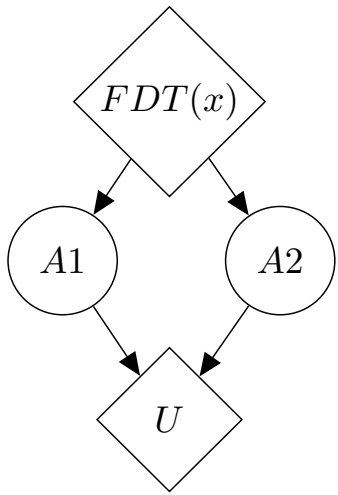

Under FDT, given that the two agents are "twins", we represent this fact by modeling the actions of both agents as dependent on the same decision function, **FDT(x)**.

When the given FDT agent acts, it intervenes at **FDT(x)**. It first tries taking **do**(FDT(x) = C) as given. Here we imagine that the values of **A1** and **A2** are determined by **FDT(x)** with probability 1. So they are both set to C, resulting in 7 utility for each agent.

The FDT procedure then tries taking **do**(FDT(x) = D) as given. **A1** and **A2** are both set to D, resulting in 4 utility for each agent. Two FDT twins in a Prisoner's Dilemma therefore cooperate and outperform two CDT twins.

**Figure 1.4.3**

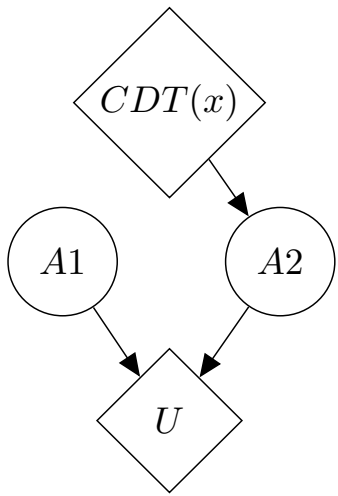

Even if CDT attempted to model the fact that both agents are dependent on the same decision function, when the agent intervenes at **A1**, it severs all incoming connections. It then leaves **A2** fixed as normal as it iterates over possible values of **A1**.

**Figure 1.4.4**

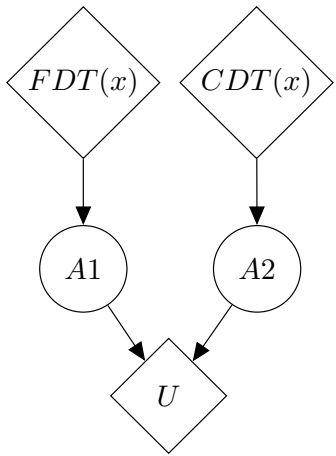

If an FDT agent is in a Prisoner's Dilemma against an agent that does not share its decision function, such as a CDT agent, it stops cooperating. As one can see in the graph, the FDT procedure intervenes at **FDT(x)**, setting it equal to C and D, while holding its opponents action fixed. This is just the same as what CDT does, so here FDT defects.

Two agents being perfect twins may seem to be a very unrealistic assumption. One response to this objection is that in the realm of AI, it might not be so unrealistic after all. If optimal agents are ever actually built and implemented in a computer system, it could be easy to generate duplicates run on other computers. Indeed, if there is only one way to make a truly optimal agent, then if we speak at all of optimal agents competing against each other, one might think the agents really ought to be twins.

Another response would be that this logic extends beyond perfect twins. If two agents merely have *similar* decision functions, that is a relevant fact that FDT can capture but CDT cannot. For example, in the Prisoner's Dilemma, if two agent's decisions are 80% correlated rather than 100%, taking do(FDT(x) = C) as given sets one's own action to C with probability 1, and the opponent's action to C with probability 0.8 and D with probability 0.2. With the utilities above, cooperating is still optimal.

FDT then cooperates only if the correlation with its opponent genuinely exists and is strong enough to make cooperating have higher expected utility. Otherwise, it defects. In that sense, it seems to strictly dominate CDT in the Prisoner's Dilemma.

**Predictive Dilemmas**

In twin dilemmas, variables are logically dependent on the same decision function because two agents are using similar decision functions. However, there is another possible source of logical dependencies among variables. We saw it before in Newcomb's Problem. If one agent makes a decision based on the decision function of another agent, the two agents' actions will be correlated, even though the agents are not using the same decision functions at all.

The predictor in Newcomb's Problem is not necessarily using the same decision function as the agent, but it makes its prediction based on its analysis of the agent's decision function. The agent's decision function is an input into the predictor's decision function. That is why the agent's action and the predictor's predictions are correlated.

This forms a general class of predictive dilemmas, in which at one agent is able to predict the behavior of another through some sort of direct access to the other agent's decision function.

We have already seen Newcomb's Problem as an example, but let us see another. In Parfit's Hitchhiker (1986), the given agent is stranded in the desert and may die without help, a risk the agent values at -$1,000,000. A man drives by and offers to help for a reward. The agent has no money on hand, so the driver asks for a $1,000 reward when they get back to town. The agent may promise this reward, but nothings binds them to it once they reach town. The driver does, however, have a decent ability to detect lies: 70%, say. If he thinks the agent is lying when they promise the reward, he will leave them in the desert rather than waste his time for nothing. What should the agent do if they get to town?

This is quite similar to Newcomb's Problem. The driver is acting as the predictor. If the driver predicts that the agent is willing to give up $1,000, they provide the agent with a $1,000,000 benefit. If the driver thinks the agent will try to take the free ride and keep the reward, the driver leaves the agent with only its $1,000. The only significant difference is that the agent gets to see the driver's prediction before it makes its final choice.

**Figure 1.4.5**

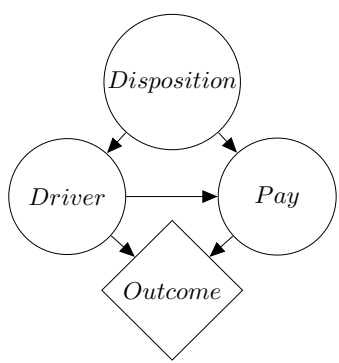

We have a graph similar to Newcomb's Problem, where the action is **Pay**, the probability that the agent will pay upon reaching town, and the state is **Driver**, the probability that the driver anticipates the agent paying. Since the agent gets to see the driver's prediction, we allow for the possibility that **Driver** influences **Pay**. The agent's **Disposition** influences their final choice and the prediction of the driver.

**Figure 1.4.6**

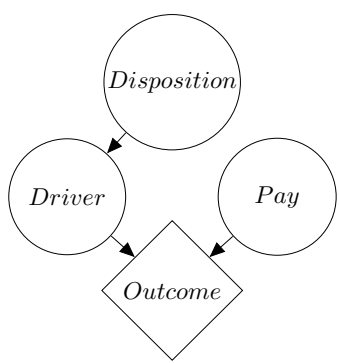

Under CDT, we intervene at the action node **Pay**, severing all incoming connections. The agent considers paying and not paying, while leaving the driver's prediction **Drive** fixed. Paying would then simply result in losing $1,000 and changing nothing else, so the agent refuses to pay.

CDT always refuses to pay. The only unfortunate thing about this is, while the agent is in the desert, it *knows* it will reason this way when it reaches town. If the agent promises the driver they will pay, the driver will likely see this as a lie and leave the agent behind. The better the predictor is, the more likely the agent is to be left for dead, but since CDT behaves as if its actions are independent of all predictions made about itself, it has no way to capture this fact. The agent's expected value on this problem is 0.7(-$1,000,000) + 0.3($0) = -$700,000.

**Figure 1.4.7**

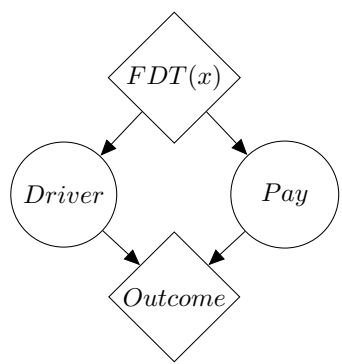

FDT models its decision function as determining its own action, as well as the prediction of the driver. When the driver sizes up the agent and searches for lies, he is seeing into the agent in some sense, seeing a piece of how the agent makes its choices.

Under FDT, we intervene at **FDT(x)**, but sever all other incoming edges from the relevant action node. Taking **do**(FDT(x) = "Pay") as given, then **Pay** = 1 and **Driver** = 0.7, an expected value of 0.7(-$1,000) + 0.3(-$1,000,000) = -$300,700 for paying.

Taking **do**(FDT(x) = "Refuse") as given, then **Pay** = 0 and **Driver** = 0.3, an expected value of 0.7(-$1,000,000) + 0.3($0) = -$700,000 for not paying. FDT therefore pays, and outperforms CDT.

This reasoning seems to work and achieves higher expected utility than CDT's reasoning. But the reasoning may seem strange when one starts to track the counterfactuals. When our FDT agent gets to town, it thinks "If I pay, I lose $1,000. But if I do not pay, I likely would have been left in the desert, which is worse than losing $1,000, so I will pay."

How could a choice made now retroactively change the past? Remember that counterfactual reasoning is just a conceptual tool and may not always make literal sense. The CDT agent's counterfactual reasoning says that its actions can vary while all predictions being made about it stay fixed, even if the predictor has direct access to the decision process the CDT agent is using. This also does not make literal sense, so the only question is, which mode of reasoning leads to higher expected utility? And the answer is FDT, because it is left in the desert much less often.

This seems significant, because it is common in game theory for opponents to model each other. If these models are accurate enough to actually capture some of an agent's decision function, a CDT agent victim will not be able to take this into account.

This may seem far afield from standard game theory reasoning, so let us see if we can look at it from a more classical viewpoint. In many ways, attempting to model another individual agent's decision function is very similar to trying to judge them based on their reputation.

Reputation is a common idea in game theory applied to games that are played repeatedly between the same opponents, or if one happens to know how their opponent played in previous games against others. This assumes we have an imperfect idea of how our opponent makes their decisions, and we use their past behavior to infer their future behavior. Every action they take reveals a bit more about their decision-making process.

One of the interesting implications of this is that it often leads to agents foregoing a potential gain in one round of a game in order to build up their reputation and receive more future benefits. This is often seen in the Iterated Prisoner's Dilemma. It is classically seen as irrational to cooperate in the Prisoner's Dilemma. But what if cooperating now can build up one's reputation so that others will be more willing to cooperate in the future? While defecting in this round would lead to more immediate gains, one can forgo these gains for an increase in future benefits. Defecting now has a reputational cost, and will make one appear less trustwor-thy in the future. Sustained cooperation can be attained in classical game theory on these grounds.

Opponents trying to model each other is very akin to this. It is an attempt to judge the opponent's "reputation", but potentially without knowledge of their past behavior. Instead, one has some limited access to their decision process *right now*. Unless the opponent is 100% opaque, then one will always have some access like this, some ability to judge what they will do in this round of the game. In classical game theory, your action in *this* round can only ever affect your *future* reputation. But under this new lens, if an agent is transparent to any degree, then their planned action *this* round can affect their reputation *now.*

So while classically one would only every cooperate now to try to receive more cooperation in the future, under FDT, it is possible to see that choosing to cooperate now will affect the decisions of opponents on this very round. So choosing to defect, even in a one-off game, can potentially have an instantaneous reputational cost, a reputational cost that is not long-term, but instead makes the opponent trust one less on this very round. The fact that our CDT agent is willing to betray the driver in town has a reputational cost back in the desert, because it makes the driver less willing to help the agent.

**Costless Commitment**

Another common idea in game theory is that of commitment. Notice that when our CDT agent is stranded in the desert, it may very well wish that it could make a binding promise to pay the driver in town, because this would convince the driver to take them. The problem arises because the promise cannot be enforced, and so the CDT agent knows it will break it. If it had a way to visibly commit itself to the action of paying later, it would do so.

It may seem strange that an agent could do better by having *fewer* options. If an agent is rational, surely having more options would always be better? The agent can always simply ignore the bad ones. But this is not how CDT works. This is even more concerning when one considers that, in realistic terms, committing to some course of action must come with a cost. It is difficult to commit, or else it would not even be a problem. One must tie their hands, stake their livelihood on this choice, whatever they have to do to stay committed. So a CDT agent must pay a *penalty* just to give itself *fewer* options if it wishes to achieve more utility.

Notice that an FDT agent does *not* have these issues. The problem with the CDT agent is that it knows it would choose not to pay once it reached town, and so a commitment is necessary to override this fact and removes its ability for choice. An FDT agent, on the other hand, freely chooses to pay, and so requires no commitment to do so. This is similar to the Iterated Prisoner's Dilemma, where an agent does not need to commit itself to cooperating, it simply sees that it will achieve higher utility in the long-run. Likewise, an FDT agent does not need to commit itself to paying up, because it honestly believes paying achieves the highest expected utility.

In general, if an FDT agent is facing an accurate predictor, it considers its decision process to be as transparent as a public commitment. When considering potential outputs of its decision function, the agent then outputs the exact same decisions as it would like to have pre-committed to, and it does so at no extra cost and without needing an opportunity to commit in ad-vance of the game beginning. In reality, this is a matter of degree. The more predictive power the opponent has, the more effectively FDT can implement costless commitment.

# 2 - Evolutionary Environment

In evolutionary game theory, to test how effective different strategies are in a game, an artificial environment is constructed. A large population of agents is created in a mathematical or computerized simulation. Different groups of agents follow different strategies. Agents are randomly paired off many times to face each other in the game, and their average utility earned is saved over time. After a while, high-performing agents reproduce, and low-performing agents die off. Over time, we analyze which strategy seems to dominate the population. We will do something similar here. The interested reader can learn more about these methods from *Game Theory Evolving: A Problem-Centered Introduction to Modeling Strategic Interaction*.

## 2.1 Competitors

Due to EDT's problems and unpopularity, the main competitors in this environment will be FDT agents and CDT agents. Most evolutionary environments consist of various agents who follow different pure- or mixed-strategies, also known as having different types, rather than representing a general decision algorithm. But our agents are still following a fixed strategy, defined by their utility-maximizing equation. It is simply a more elaborate strategy. We still use simpler agents on occasion, who do follow simple strategies, depending on the game.

We will have a large population, and we will randomly pair agents together to face off in a game. So we will see CDT vs. CDT, FDT vs. CDT, and FDT vs. FDT. This process will be repeated many times, and agents will then be added and removed from the population in pro-portion to how much utility they earned in their games, and we will see which agents dominate the population over time. We will apply this analysis to various games.

We hope this will dissolve some philosophical questions of what it means to be rational. FDT agents often freely choose to perform actions that are seen as subgame imperfect. This means they choose actions that have no apparent present incentive, such as paying up in Parfit's Hitchhiker or one-boxing in Newcomb's Problem.

Proponents argue that being the kind of agent who is willing to pay is beneficial, because one is then less likely to end up in disadvantageous scenarios in the first place. Others disagree, and would question what sense it makes to minimize the probability that you end up in a bad scenario *once you are already in it*. An evolutionary environment has the potential to demonstrate the FDT proponents' argument: if agents who are willing to pay spread over time, this directly shows the advantage we are discussing.

## 2.2 Transparency

This environment will be partially-observable and deterministic. This means our agents will not be able to fully observe everything about their opponent. They do not have full knowledge of how their opponent will make their choices, but they will have a chance to analyze each other and receive imperfect information. Determinism we have already covered, but it simply means that all of our agents' decisions are following a set process, whatever it may be.

Classical evolutionary environments, as we said, are often filled with agents of many types. In some setups, agents are able to discover each other's types with limited ability, such as the Trust in Networks problem (Herbert 2009). This is essentially how we will model our agents trying to predict each other. The signal will imperfectly identify either the opponent's type or the opponent's final action, as is appropriate to the particular game at hand.

So we assume that our participants are not perfectly opaque. They give away something of their plans to the readily observant. This is again very similar to Parfit's Hitchhiker. The driver has some ability to see what the agent's planned future decision is. He does not necessarily know how they will make that choice. In Newcomb's Problem, we usually imagine that the predictor *is* able to analyze the agent's decision function in detail. We allow for both of these possibilities.

Herein lies the important differences between CDT and FDT once again. A CDT agent always assumes that its actions are independent of any predictions made about itself. The opponent's prediction occurs before the agent's action, so how could their action causally affect the prediction made about their current behavior? A CDT agent imagines that the opponent selects some prediction from a random distribution, and once this is set in stone, the agent may act however they like with no implications on the prediction made about itself.

An FDT agent assumes quite the opposite. The prediction being made is based on the opponent sizing the agent up, analyzing them, and predicting what they will do on this round. If the opponent can predict the agent's final decision with 70% accuracy, then when the agent imagines taking different actions, they counterfactually imagine a different prediction being made. Which of these lines of reasoning leads to higher expected utility? This we will see.

That covers the case of predictive games: with probability $p$, one agent can predict the other agent's action and take that into account. In the case of twin games, $p$ could instead represent the probability that each agent outputs the same action as their opponent, if they are in the same set up. This $p$ represents how similar the two "twin" agents really are. Or, if the agents are exact twins, $p$ could represent how certain the agents are of this fact, or the twins could both be executing a mixed strategy at the fixed rate of $p$.

## 2.3 Realism

How realistic is this environment that we have constructed? Is this just a fanciful idea, or is it grounded in reality? In classical game theory, it is generally assumed that one cannot extract any extra information about an opponent beyond just knowing their preferences, and perhaps their past behavior. But when trying to bring this reasoning into the real world, it seems clear that if our agents are human, they may be able to *see* each other. When a person knows their opponent's preferences but is still uncertain what their behavior will be, being able to see them may give away a good deal of information. Do they look nervous? Confident? Neutral?

We argue that there are very good reasons to believe that humans really do exist in something like the environment we have described. If we did not, there would be nothing to the game of poker beyond studying the cards. Trying to predict an opponent's behavior from their face or mannerisms would be useless if we did not assume that humans inherently let something of their plans and decision process show on their face.

Some biologists and anthropologists believe that this was a key driver in the evolution of human intelligence: the ability to detect and conceal lies from others, to build up a theory of mind to outwit competitors. This is referred to as the Machiavellian intelligence hypothesis (de Waal 1982). Humans are semi-transparent. Their intentions are partially visible to others. People can simply *appear* less trustworthy, even without reputational knowledge, and it is evolutionarily ingrained in us to recognize certain behaviors as suspicious, such as aggressive or fidgety body language.

One way to get around this is to actually *be* trustworthy, which may be why such social norms have evolved in humans. One could say that commitment to a certain amount of trustworthiness has evolved *into* humans, and now this is simply the way we reason. We do not always need to take some action to commit ourselves to behaving trustworthy. Instead, norms such as honor are simply inherent in how we make choices, or ingrained by culture, and something like the evolutionary environment we have described may be responsible for this. We hope at least that it does not appear to be a ridiculous notion.

Moving away from humans, we consider our other previously-mentioned agent type: artificial intelligence. An AI could one day possess a fully-described and fully-general decision algorithm for problem solving, specified in computer code. If two such agents faced off in a game, and they had some limited ability to access each other's source code, this would make our evolutionary environment more than plausible. The agents would have a concrete way to predict their opponents. They can analyze the code, run simulations, etc. In the extreme, full access to an opponent's entire decision algorithm could lead to perfect predictive power, right here in the real world. We would like to be able to outline an agent robust enough to handle this possibility.

## 2.4 Repopulation Mechanics

For any particular game, we will start off with a large population of agents (say 10,000). We may initially evenly split this population between the agents we are using, although we can experiment with the initial conditions to see what happens. We randomly pair agents from this population to face-off in the game for a number of rounds (say 100). We track the utilities they earn during this time.

To repopulate, we set a certain birth/death rate (say 1%). We then randomly select 1% of agents to copy/reproduce, weighted by their earned utility.[4] We randomly select 1% of the agents to kill off, inversely weighted by their earned utility. We thereby randomly eliminate low-utility agents and spread high-utility agents. We also have a smaller mutation rate (say 0.1%). We uniformly randomly select 0.1% of the agents in the population and set their type to a uniformly random type. We can repeat this for many generations, dependent on the game. The specific parameters in each problem are chosen to be just large enough to settle on a clear winner. We track how the populations rates change over time given different initial conditions, testing the behavior of the agents in the game.

---

[4] All utility values will be positive to make this process simple.

# 3 - Evolutionary Games

We consider three games in the manner previously described. All relevant code can be found here: shorturl.at/ty678.

## 3.1 The Prisoner's Dilemma

We will start with a simple and specific case of the Prisoner's Dilemma, and gradually generalize it. Take a Prisoner's Dilemma with the following payoffs:

|  |  | Agent 2 | |
|---|---|---|---|
|  |  | C | D |
| Agent 1 | C | 7, 7 | 1, 10 |
|  | D | 10, 1 | 4, 4 |

We have a large population of players ranging over three types: Defectors, who defect no matter what, Cooperators, who cooperate no matter what, and FDT agents, whose strategy we will cover. We will initially assume an equal proportion of each agent (i.e. each type makes up 1/3 of the population), and we will randomly select pairs of agents from this population and have them play in a one-off Prisoner's Dilemma.

Despite the simplicity of Defectors, we can model them as CDT agents. A CDT agent always defects in a one-off Prisoner's Dilemma, regardless of their opponent's strategy or type. So we can imagine that the Defectors run the full CDT graphical algorithm to determine their choice; it just so happens to always settle on defection, the prescription of classical game theory. Cooperators are here largely to demonstrate that FDT agents do not simply cooperate with any agent who is "friendly", and to help contrast FDT from the simplicity of the Cooperator.

Our agents will have a chance to analyze each other before they choose their actions. Agents receive a signal $S$ from their opponent which estimates their opponent's type. This signal is accurate with a probability greater than chance. We will for now say it is accurate with probability $p = 0.9$. Type 1 ($S = 1$) will correspond to Defectors, type 2 ($S = 2$) will correspond to Cooperators, and type 3 ($S = 3$) will correspond to FDT agents.

That is to say, a Defector (or any other type) sends a signal identifying themselves as a Defector with probability 0.9. The rest of the probability mass is allocated equally among the other options, so a Defector has a 0.05 probability of sending a Cooperator or FDT signal.

Let us analyze what will occur in the first generation of agents. Our Cooperators are quite simple and make no use of the signal. Interestingly, the same is true for our CDT agents. Although they are aware of the signal, their action cannot causally affect a previously sent signal, and they defect regardless of their opponent's type. So we only really need to analyze how this signal affects the FDT agents' strategy. Let us go over all the signals FDT might get and what each would mean, from the perspective of said agent.

If $S = 3$, this signals that the opponent is another FDT agent. To figure out exactly how likely this is to actually be the case, we must take into account the base rate of FDT agents in the population, as well as the strength of the signal, and use Bayes' rule to calculate the posterior probability.

We use the odds form of Bayes' rule. The odds for an agent's type are represented by a three-dimensional vector. The first entry is proportional to the probability mass assigned to the agent being type 1, and so on. To calculate the odds after making an observation, we multiply the prior odds ratio by the likelihood ratio of the observation. The likelihood ratio is another three-dimensional vector, where the first entry is proportional to the probability mass assigned to making the given observation if the agent is of type 1, and so on.

Initially, all agents are equally common, so the prior odds on the opponent's type are (1 : 1 : 1). The likelihood ratio of getting the signal $S = 3$ is (5 : 5 : 90). Multiplied together, the odds ratio for the agent's type is (5 : 5 : 90).

Converting these odds into probabilities gives us $P(\text{Type} = 3 | S = 3) = 0.9$, and a probability of 0.05 for each other type. Since the base rates start off equal, our probabilities come directly from the signal, but this will not be true in general as the population shifts. If FDT agents were very rare in the population, then even given this signal, we may conclude that the opponent is likely of a different type.

If the opponent is an FDT agent, they are very likely to make the same decision as the given agent. In fact, if they received the same signal, we assume that they will definitely make the same decision. When FDT receives "$S = 3$" as an input, it must always produce the same output, assuming the given population rates are also the same. Upon receiving a signal of type 3, we will consider the expected utility of setting $FDT(S = 3)$ equal to cooperation, followed by defection, and settle on whichever has higher expected utility, calculated like so:

$$EU\big(FDT(S=3)=C\big) = 0.05(1) + 0.05(7) + 0.9 \cdot \Big(0.9 \cdot U\big(C, FDT(S=3)\big)$$
$$+ 0.05 \cdot U\big(C, FDT(S=2)\big) + 0.05 \cdot U\big(C, FDT(S=1)\big)\Big).$$

This equation should be read as follows. The expected utility of setting FDT(S = 3) equal to cooperate is equal to a 5% chance of accidentally meeting a Defector and receiving 1 utility, plus a 5% chance of meeting a Cooperator and receiving 7 utility, plus a 90% chance of meeting another FDT agent. This FDT opponent has a 90% chance of receiving the proper signal that the given agent is FDT, in which case we receive whatever utility occurs when the agent cooperates and the opponent execute FDT(S = 3). There is a 5% chance that the FDT opponent incorrectly receives the signal that the agent is a Cooperator, and a 5% chance that they are a Defector. In those cases, we receive whatever utility occurs when the agent cooperates and the opponent executes either FDT(S = 2) or FDT(S = 1), respectively.

Given that we are setting FDT(S = 3) equal to cooperate, we can plug that value into the equation. U(C, FDT(S = 3)) then becomes U(C, C), which is 7. If we also collect the constants, the equation simplifies to:

$$EU\big(FDT(S=3)=C\big) = 6.07 + 0.045\Big(U\big(C, FDT(S=2)\big) + U\big(C, FDT(S=1)\big)\Big).$$

We repeat this exercise with Defection. The initial equation is:

$$EU\big(FDT(S=3)=D\big) = 0.05(4) + 0.05(10) + 0.9 \cdot \Big(0.9 \cdot U\big(D, FDT(S=3)\big)$$
$$+ 0.05 \cdot U\big(D, FDT(S=2)\big) + 0.05 \cdot U\big(D, FDT(S=1)\big)\Big).$$

This time, we plug in D for FDT(S= 3). U(D, FDT(S = 3)) becomes U(D, D), which is 4. Collecting constants, we have:

$$EU\big(FDT(S=3)=D\big) = 3.94 + 0.045\Big(U\big(D, FDT(S=2)\big) + U\big(D, FDT(S=1)\big)\Big).$$

If we try plugging in all possible combinations of values for FDT(S = 2) and FDT(S = 1), we will see that, regardless of what we choose, EU(FDT(S = 3) = C) > EU(FDT(S = 3) = D). That is, the expected utility of FDT cooperating given the signal S = 3 is always greater than the expected utility of defecting, given the numbers we have used in this setup. Therefore, we output FDT(S = 3) = C.

This process can be repeated for the cases of S = 2 and S = 1. Since the prior population base rates were all equal, and $p$ is the same for each type of agent, the probabilities will be symmetric. For example, if S = 2, the probability of the opponent being a Cooperator is 0.9, and the probability is 0.05 for FDT and Defector. The setup is very similar, and it turns out that FDT will defect in both cases, S = 2 and S = 1.

This is intuitively obvious. If the opponent will always defect or always cooperate, the agent always stands to gain by defecting. Graphically speaking, we would say that the agent's and opponent's decision functions are not logically dependent, unlike the case of FDT vs. FDT. If the agent is 90% sure that this is the case, then given our chosen utilities, it will defect. The math bears all this out, but we omit it since it is so similar to the previous process, and produces unsurprising results.

Let us now have our agents go head to head. What is the expected utility of being an FDT agent on a particular round in the first generation of this game? If we label the description of the scenario $x$, then:

$$EU\big(FDT(x)\big) = \frac{1}{3}\big(0.95(4) + 0.05(1)\big) + \frac{1}{3}\big(0.95(10) + 0.05(7)\big)$$

$$+ \frac{1}{3}\big(0.81(7) + 0.09(10) + 0.09(1) + 0.01(4)\big) = 6.8$$

This equation should be read as follows. There is a 1/3 chance the agent runs into a Defector. In this case, the agent has a 90% chance of recognizing the opponent as such, plus a 5% chance of mistakenly identifying them as a Cooperator. In either case, the agent defects and gets 4 utility. Alternatively, the agent could mistakenly identify the opponent as an FDT agent (5% chance), and decide to cooperate, receiving 1 utility.

There is a 1/3 chance that the agent runs into a Cooperator. The agent has a 95% chance of identifying the opponent as a Cooperator or Defector and deciding to defect, receiving 10 utility. The agent also has a 5% chance of mistakenly identifying the opponent as an FDT agent and cooperating, receiving 7 utility.

Finally, there is a 1/3 chance the agent runs into another FDT opponent. There is an 81% chance that both agents receive the correct signal and both cooperate, receiving 7 utility each. There is a 9% chance the agent receives the incorrect signal and defects, while the opponent receives the correct signal and cooperates, resulting in 10 utility for the agent. There is a 9% chance of the reverse, resulting in 1 utility for the agent. Finally, there is a 1% chance that both agents receive the incorrect signal and defect, resulting in 4 utility each. Summing all these together results in an expected utility of 6.8.

The corresponding equations for CDT and Cooperators are:

$$EU\big(CDT(x)\big) = \frac{1}{3}(4) + \frac{1}{3}(10) + \frac{1}{3}\big(0.95(4) + 0.05(10)\big) = 6.1$$

$$EU\big(COOP(x)\big) = \frac{1}{3}(1) + \frac{1}{3}(7) + \frac{1}{3}\big(0.95(1) + 0.05(7)\big) = 3.1$$

Overall, then, our FDT agent has the highest expected utility. It gains this advantage from its willingness to cooperate when prudent, its observation of its environment, and its understanding of determinism and counterfactuals. We should thereby expect FDT to spread in this population. This is only a specific case, however. Does it hold in general? And what happens as we update the population?

The general parameters for this set up are the initial proportions of each agent, the payoffs of the Prisoner's Dilemma, and the strength of the signal. The FDT algorithm is certainly dependent on these, so it will not invariably output "defect given signal 1 or 2, and cooperate given signal 3". If the signal is too weak to identify FDT agents reliably, or if the negative payoff from losing is too high, then FDT will defect even given signal 3.

Notice, however, that FDT only ever chooses to cooperate if the expected utility of doing so is higher than defecting. This is what defines its algorithm. So whenever the parameters are such that ever choosing to cooperate would be a bad strategy, FDT ceases to do it, and instead behaves indistinguishably from a Defector/CDT agent. FDT can only ever do better than CDT, never worse. We could say FDT *dominates* CDT in this set up.

Fully solving and displaying the results of this dynamical system would be challenging. To tackle this problem, we will have a computer do the work instead. With a relatively simple Java program, we can model this problem in the manner described in the introduction to our evolutionary environment. To empirically demonstrate the behavior of FDT, we test it with several different initial conditions. We will use a birth/death rate of 1% and a mutation rate of 0.1% throughout all of our evolutionary games.

To start off, we use the parameters from our example problem above and see what happens beyond just the first generation. The simulation is run on a population of 10,000 agents over 750 generations. In each generation, agents are randomly paired 100 times and face-off in a Prisoner's Dilemma as described. Their scores are saved and summed, and the population is repopulated at the end of each generation as previously described. Here is the population after certain generations:

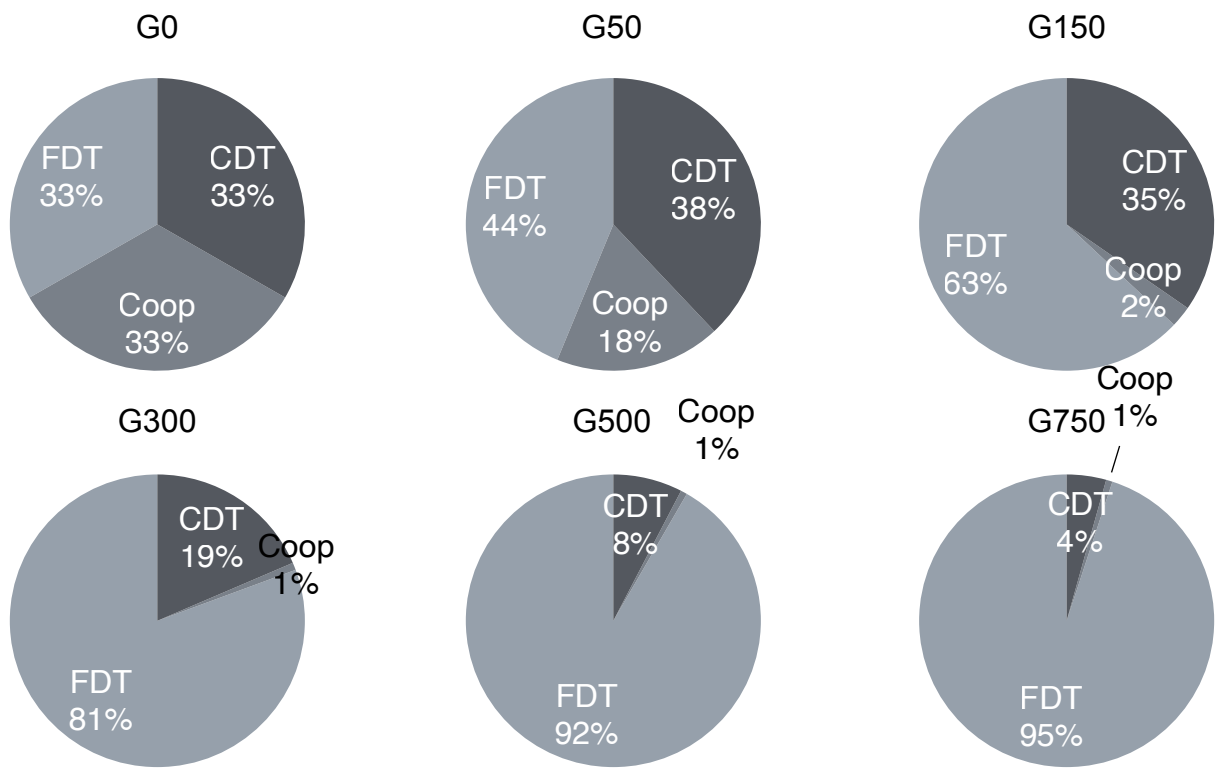

The conditions chosen may seem to unfairly skew in favor of FDT, however. We can try a few ways to change this: weaken the signal, vary the payoffs, and change the initial population. Let us look at each of these cases.

Obviously, we can set the signal below any useful value. If $p < 0.5$, it means little. Instead, let us set $p$ around 0.5 and higher to see what happens. Each run here was for 2000 generations.

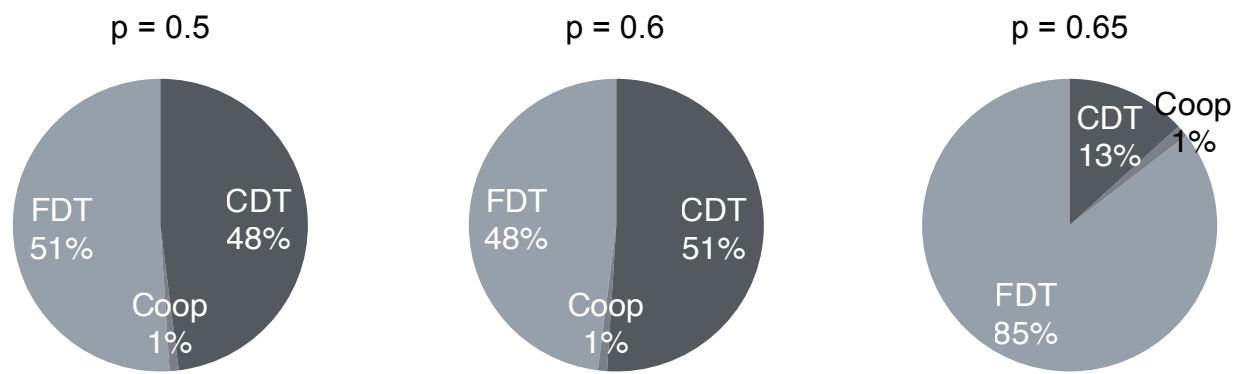

For $p = 0.6$ and below, the signal is not strong enough for FDT to see cooperating as ever beneficial, so it always defects, and thereby spreads in the population at the same rate as CDT. Right around $p = 0.65$ is a strong enough signal for FDT to eventually gain dominance. The exact signal strength necessary is of course an artifact of the selected parameters.

Resetting our parameters, we can now vary the payoffs. To see how FDT generalizes to various payoffs, we can try randomly setting the payoffs, subject to the constraint that they still constitute a Prisoner's Dilemma. We will set the payoffs to be random integers from 1 to 1000. We do this six separate times to test different payoff combinations. Each run below is presented at the end of 1000 generations.

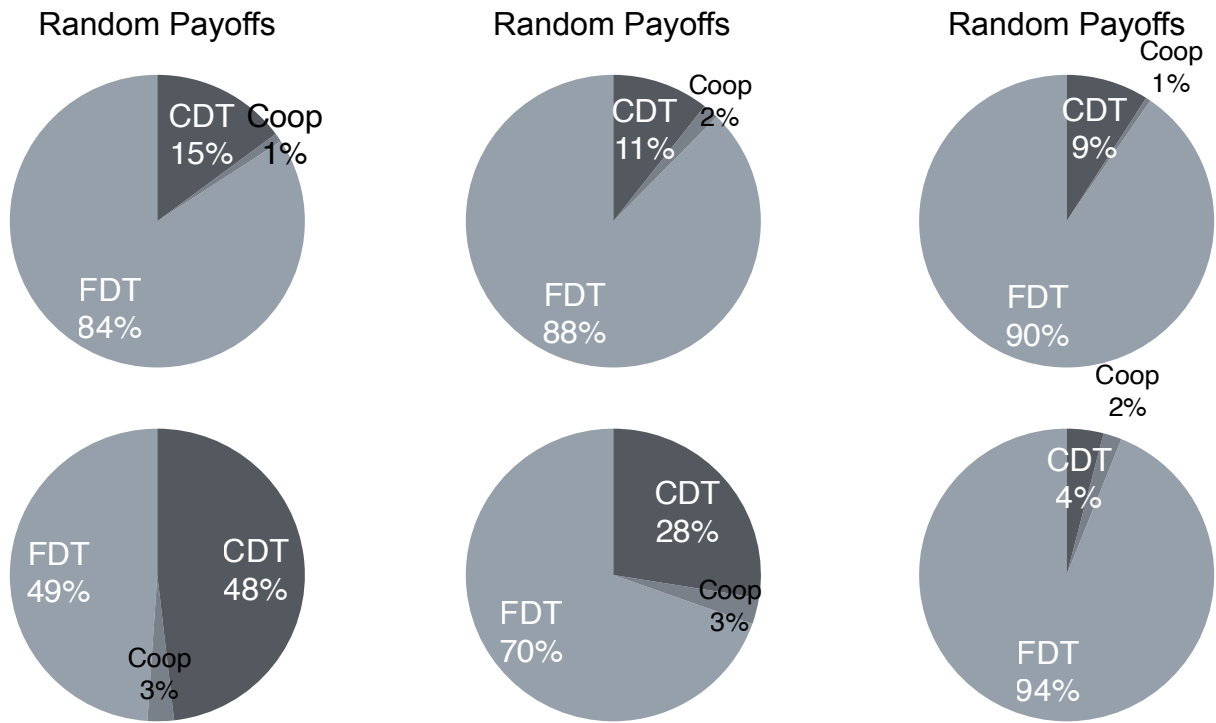

These were our very first six runs, so FDT seems to generalize very well on random payoffs.

Finally, we will reset our parameters and try to give CDT a big advantage: boost its initial population size. CDT will start off dominant, making it especially difficult for FDT agents to ever find other FDT agents. We will run this for 1500 generations and take a look every 300.

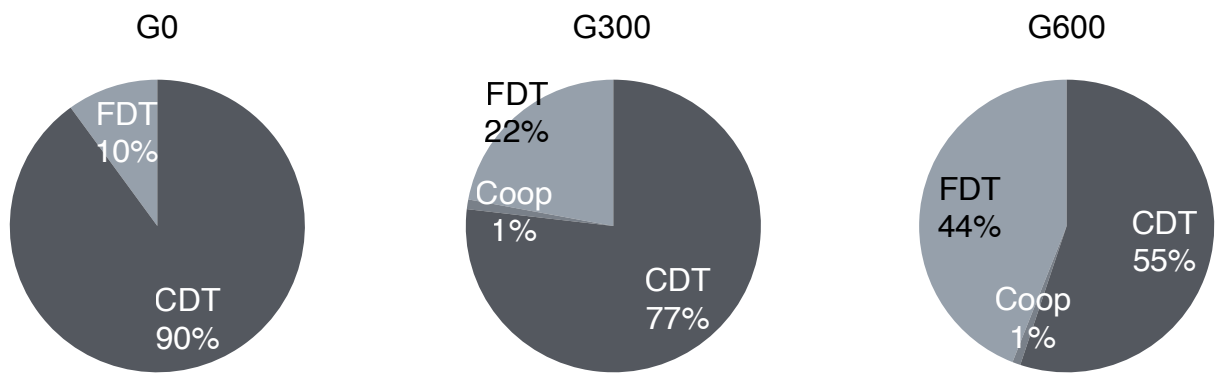

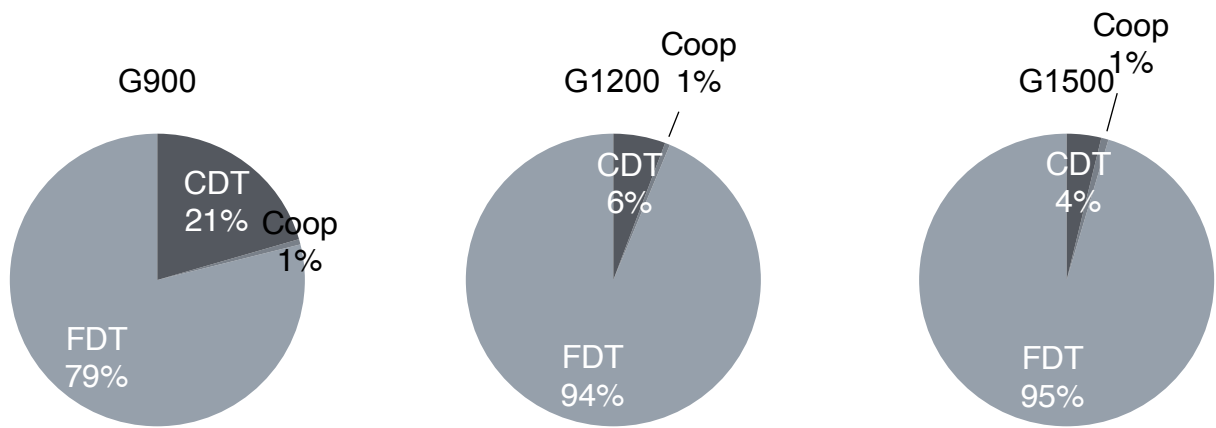

This clearly took longer, but FDT still dominates in the end. The takeaway here seems to be that CDT is vulnerable to invasion, while FDT is not. Indeed, setting the mutation rate to zero removes the effect, and the ratio remains around 9:1. But with mutations, FDT can invade the CDT majority. Right around when the two populations are equal is when FDT agents start cooperating with each other and pushing toward totality.

All this seems to verify what we would expect from the math and theory. FDT never underperforms CDT, FDT is invulnerable to invasion while CDT is not, and FDT in general tends to dominate in the environment. FDT is surprisingly robust to adverse conditions. The only necessary assumptions are that the agents have a modestly-above-chance ability to read each other, and that FDT agents acknowledge that their decisions come from the same decision function. Both of these are open to the CDT agent as well. They know just as well as FDT what the probabilities are for their opponent's type, they simply ignore the correlation between their decision algorithm and their twins', leading to their failure.

## 3.2 Transparent Newcomb Problem

We have already covered Newcomb's Problem, but now we modify it slightly. Instead of the predictor presenting the agent with one opaque box and one transparent box, both boxes are now transparent. Box A always contains $1,000, and Box B contains $1,000,000 if and only if the predictor believes the agent would take just Box B upon seeing both boxes full. Otherwise Box B is empty. This is the Transparent Newcomb Problem.

If a given agent encounters this game and is presented with two full boxes, should it take both, or just the $1,000,000? The only change from the original Newcomb's Problem is that the agent now gets to see the prediction of the predictor before the agent makes their final choice. This may sound familiar, because that is also the only difference between Newcomb's Problem and Parfit's Hitchhiker.

The two games are in fact equivalent. When the agent sees both boxes full, it is in the same scenario as the agent who has made it to town. Both seem to have a guaranteed value of $1,000,000 and must simply choose between losing or keeping $1,000. The problem has the same graphical structure as that of Figures 1.4.5 - 1.4.7.

Therefore, FDT will one-box and CDT will two-box, since these are analogous to paying up and refusing to pay, respectively. We take CDT's behavior to be obvious, but FDT's choice may still seem strange. The agent can see the $1,000 sitting in the box, so why would it leave it there?

FDT's reasoning is the same as before, assuming the predictor is reasonably accurate. When it imagines two-boxing, it imagines the prediction of the predictor changing, leaving the agent with just $1,000. Being able to see into the box does not change this.

In fact, although FDT's behavior may seem stranger here, it actually performs better than in Newcomb's Problem. This is because the agent will no longer ever end up with nothing. If the predictor ever makes the wrong prediction, the agent will see Box B empty, and can freely choose to take the $1,000. The predictor's prediction here is only dependent on what the agent would do if it saw both boxes full. As we would hope from an optimal theory of rationality, FDT performs better when it has more information.

Leaving the $1,000 behind may still feel intuitively wrong. But we would rather see the issue settled empirically than by philosophical argument. We therefore make our next evolutionary environment to model the Transparent Newcomb Problem and see which agents perform better in general. Since they are equivalent, the reader may imagine this model representing Parfit's Hitchhiker. This could be beneficial if that game seems more realistic.

We have essentially solved a specific case of this game previously, so let us set up the problem generally. We have something similar to our signal strength before, which is now the accuracy of the predictor. The predictor knows with probability *p* what action the given agent will choose. Agents give off this signal to the predictor before making their choice. We also have the payoffs of Box A and Box B as parameters. B is clearly meant to be greater than A, but we may vary by how much. We will call these the High and Low rewards for now.What does FDT do in general? If the agent sees only Box A full, it will take it and receive the Low reward. If it sees both boxes full, its behavior depends on the value of *p*, High, and Low.

**Figure 3.2.1**

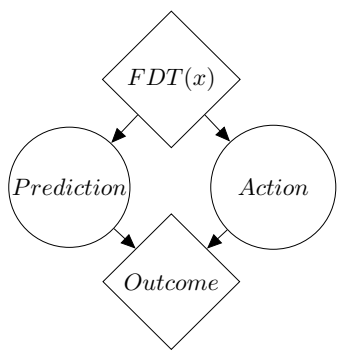

Taking **do**(FDT(x) = "One-Box") as given, the predictor will anticipate this with probability *p*, in which case they fill both boxes, and FDT receives High. Otherwise, the predictor makes an error and leaves Box B empty, in which case FDT takes Low, for an expected utility of $p \cdot High + (1-p) \cdot Low$.

Taking **do**(FDT(x) = "Two-Box") as given, the predictor will anticipate this with probability *p*, in which case they leave Box B empty, and FDT is left with Low. Otherwise, the predictor makes an error and fills Box B, in which case FDT receives High + Low, for an expected utility of $(1-p) \cdot (High + Low) + p \cdot Low$.

Typically, if *p* is better than chance (above 0.5) and High and Low are reasonably far apart, FDT will one-box. Therefore, FDT's expected utility is either $p \cdot High + (1-p) \cdot Low$ or $(1-p) \cdot (High + Low) + p \cdot Low$, whichever is higher. CDT, on the other hand, always two-boxes. The predictor anticipates this with probability *p*, in which case it leaves Box B empty. Otherwise, it fills both boxes. CDT's expected utility is then $(1-p) \cdot (High + Low) + p \cdot Low$.

If $p \cdot High + (1-p) \cdot Low > (1-p) \cdot (High + Low) + p \cdot Low$, then FDT will have a higher expected utility than CDT, perform better in general, and spread throughout the population. Otherwise, both will perform equally well.

Starting with a population of 3,000 agents, evenly split between CDT and FDT, we run this simulation for 100 generations, with 100 rounds in each generation. Here, agents are not paired off. Rather, each will enter the Transparent Newcomb Problem against a predictor. The high reward is 10,000, the low reward is 1,000, and the predictor has 99% accuracy. Agents are reproduced based on how they perform over the 100 rounds each generation. Here is one run after after 100 generations.

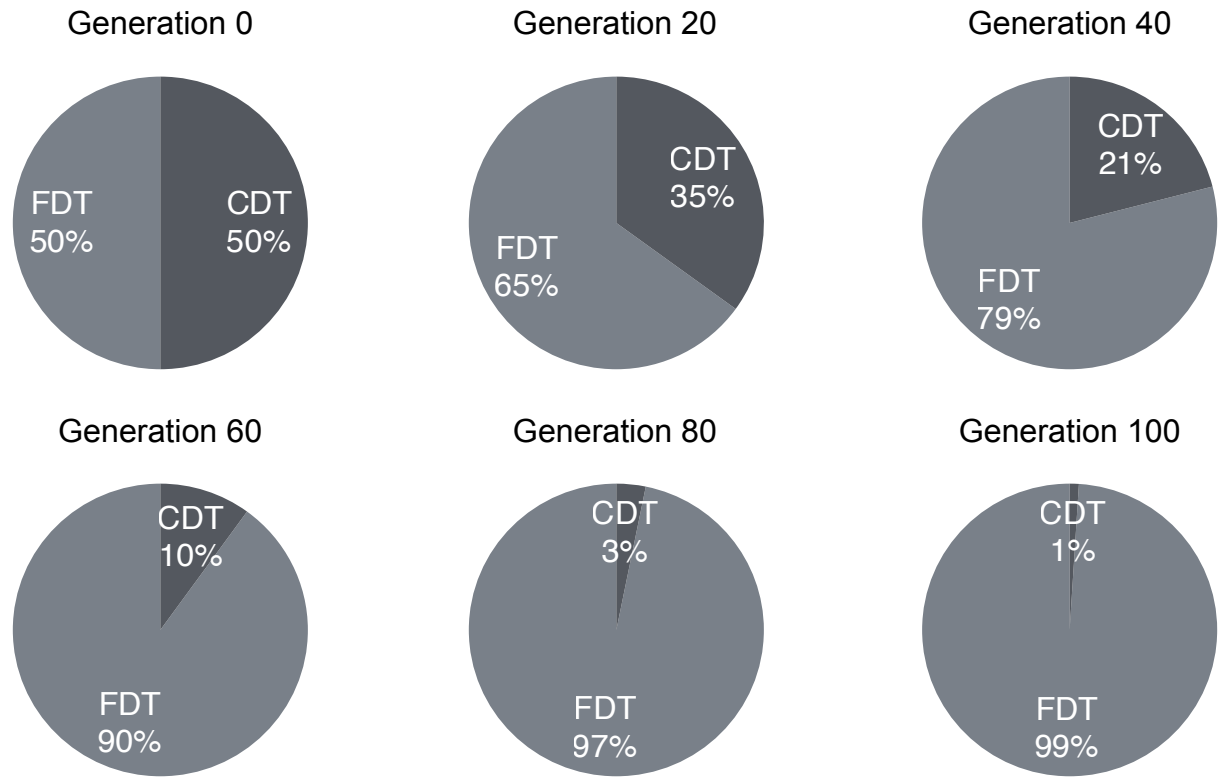

FDT's convergence is so fast that even initially setting CDT to be 100% of the population does not affect where the population is 200 generations later. How general is FDT's strength on this problem? Supporters of causal decision theory sometimes say that two-boxing only makes sense if the predictor has perfect accuracy. This makes little sense in our utility equation. In this simulation, the only assumptions are that the predictor's claimed abilities are accurate, and that CDT considers all predictions made about it in the past to be causally independent of its current behavior. How robust is FDT's success in this environment to changes in the parameters?

We will randomize the payoffs with positive utility values from a uniform distribution from 1 to 1,000,000. We ensure that the high reward is greater than the low reward. We allow a wide range of values, since this is normal in Newcomb's Problem, and they are positive so that CDT is always tempted to two-box. We also randomize the prediction accuracy, uniformly selected to be greater than 0.5 and less than 1. Here, the only assumption is that the predictor's abilities are better than total random chance. Otherwise, the parameters are the same as before. Here are the first six separate runs at the end of 500 generations.

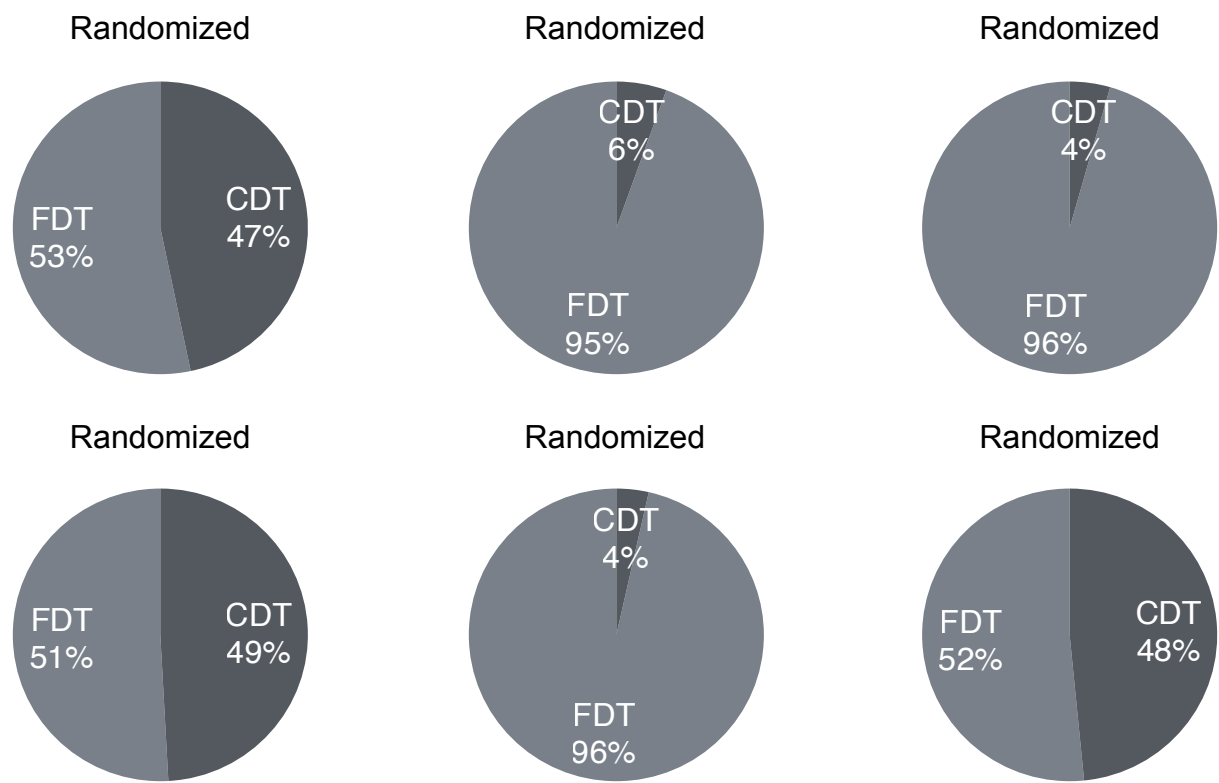

For half of these, FDT quickly dominated the population. The lowest prediction accuracy for those was $p = 0.689$. The closest the the utility values ever got was within about 58% of each other.

The point here is to stress the general applicability of Newcomb's Problem (or Parfit's Hitchhiker) to real-life situations. If people treat others modestly better for being trustworthy, and if people have modestly better-than-chance ability to assess the trustworthiness of others, then it pays to be cooperative. We in fact claim that this describes a good deal of human interaction.

## 3.3 Keynesian Beauty Contest

Our final evolutionary game is a version of the Keynesian Beauty Contest. Imagine a contest in which entrants are asked to choose which model from a group is the most attractive. Those who guess correctly win a prize. But how do we know who picked "correctly"? Presumably based on what others also think is attractive. So we will say that whoever selects the most popular model (i.e., the model most selected by others in the same contest) will win a prize.

A naive participant may simply select the model they find the most attractive. But is this actually the optimal strategy? Surely one should pick the model that they expect *others* to find attractive, not just the model they find personally attractive. But then, what if everyone else thought that as well? Then one should select the model that they expect everyone else will think the majority finds attractive… and so on. This is the Keynesian Beauty Contest (Keynes 1936).

There are many variants. We will explore the following. Participants are asked to select a real number from 0 to 100. They are trying to guess 2/3 of the average of all participants' guesses. On the first pass, if one assumes the other guesses are random, then they would tend to aver-age around 50. So one should guess 2/3 of 50, 33.33 But if everyone realizes this, then the average guess will be 33.33, so one really ought to guess 22.22… and so on. What should a participant do?

The equilibrium answer is typically taken to be 0. If all participants guess 0, none of them have any incentive to change their guess. And if the average is not 0, all participants have an incentive to guess lower than the average. This experiment has been done with human participants. One was done through a Danish newspaper with 19,196 participants and a prize of approximately 5000 Danish kroner (approximately $550). The winning guess was 21.6, so the winner performed the 2/3 iteration about twice (Politiken 2005).

It is interesting to note how closely functional decision theory follows common-sense reasoning on this problem. "But if I think that, they will think it too…". This is FDT in a nutshell. Going back to the original beauty contest, imagine that there are three models, a redhead, a brunette, and a blonde. Perhaps the majority settles on the blonde, not because any of them find blondes particularly more attractive, but because they all know blonde is often culturally portrayed as the most attractive hair color.

Under the lens of FDT, we could say that the participants share a culture that influences their decisions, and they all have common knowledge of that culture. Therefore if one participant remembers that blondes are often portrayed as the most attractive, they can bet on others remembering it too. Their remembering it does not *cause* others to remember it; it is simply a logical inference.

Now CDT could perhaps manage this by analyzing the entire culture and causally reasoning out the most likely choice of the other participants. But that is a much more difficult task, and it does not seem to be what most people do. What occurs first is their *own* thoughts, then the inference that others probably thought it as well. This chain of events is set off by their own thoughts, not by prior reasoning about the culture. Further, CDT's attempt here only works if it assumes other agents are selecting based on their cultural backgrounds. If all of them are using CDT, it gets them nowhere, since CDT treats its own actions as independent of its copies'. CDT does not provide a clear solution to these kinds of problems.

For the case of the 2/3 guessing game, common-sense reasoning also follows FDT's logic, which could be something like the following. If the given agent guesses 33.33, that is 2/3 of the average of random guesses. But other participants probably will not guess randomly, they will realize what the agent realized and guess 33.33. So the agent could guess 2/3 of that, 22.22. But what if the others think of that too? Then the agent should guess 14.81. If things keep going like this, they should simply guess 0.

But will the others really think that many levels deep? Given what level the agent thinks at, what level should they expect others to think at? Maybe if the agent guesses 0, they could expect a few others to guess 0 as well, but not everybody. Only a few people will match their thinking exactly. Others will be more dissimilar. If the agent knew how similar or different everyone else will think from themself, they might be able to make a good guess. The agent wants to know how the decision functions of the other participants are correlated with their own.

CDT cannot infer from its own level of thinking what levels of thinking are likely to occur to others, even if they are using the same or a similar decision process. Or at the very least, it considers the final actions of others to be independent of its own. So if one CDT agent settles on thinking two levels deep, it cannot infer anything about how many levels deep other CDT agents will choose to think. Which makes it difficult to determine what guess to make, even if they know how similar the other agents are to themself. Human reasoning seems to follow FDT more closely than CDT here.

If we set this up in an evolutionary environment, we can have the interesting quality that the population itself will be the environment in which the agents are playing. That is, the game is meant to be played in a population of agents of various types, so we will have that population be the group of repopulating agents itself.

We will have three types of agents, Random, CDT, and FDT. Random agents simply select a uniform random value from 0 to 100. A CDT agent knows the rates of Random, CDT, and FDT agents in the population. They can reason out what Random and FDT agents will do. But a CDT agent assumes its guess is independent of the guesses of all the other agents, including other CDT agents. There is no clear way for CDT to predict the actions of other CDT agents without simply executing their decision algorithm. And once a CDT agents has done this, it considers the result it gets to be independent of the results of other CDT agents. A CDT agent believes it can vary its own action while the actions of other CDT agents remain constant. We will therefore have our CDT agents be agnostic about the other CDT agents' actions, and only use the Random and FDT agents to predict the average.

An FDT agent knows the rates of Random, CDT, and FDT agents in the population. They can reason out what Random and CDT agents will do, and when they execute their own decision algorithm, they will assume that all other FDT agents will come to the same conclusion. They will then factor this into their guess, and settle on the guess that is consistent assuming that all other FDT agents make the same guess. Since we really will design our FDT agents like this, the assumption is valid.

The utility earned by agents in this game will be the inverse of their error, to maintain positive utility values for the ease of reproducing agents proportional to their earned utility. That is, $U(Guess) = 1 \div |2/3 \cdot Avg - Guess|$. We cap this value at 1000, so an error of 0.001 or lower is treated as perfect, and we keep the utility values bounded. The largest possible error is 100, so the utility values range from [0.01, 1000].

Let us take a look at what round one would look like. Assuming we have a population with 1/3 Random, 1/3 CDT, and 1/3 FDT, what would we expect to happen?

CDT wants to guess 2/3 of the weighted average of the guesses of Random and FDT. Since Random and FDT are initially equal in the population, and assuming that Random's average guess is 50, we have:

$$CDT(x) = \frac{2}{3}\left(\frac{1}{2} \cdot 50 + \frac{1}{2} \cdot FDT(x)\right).$$

FDT wants to guess 2/3 of the average of the guesses of Random, CDT, and FDT. We have:

$$FDT(x) = \frac{2}{3}\left(\frac{1}{3} \cdot 50 + \frac{1}{3} \cdot CDT(x) + \frac{1}{3} \cdot FDT(x)\right).$$

We have two equations with two unknowns, which can be solved simultaneously without much trouble to get CDT(x) = 23.68 and FDT(x) = 21.05. In the previous games of the Prisoners Dilemma and Newcomb's Problem, CDT and FDT could often output the same answer. Here we see that this is very unlikely.

The lesson is, even when FDT might seem unimportant since it outputs the same action as CDT, it is often calculating a different expected utility under the hood. Once we shift to a game with a continuous action, we can see this immediately. The only question left to be asked is if FDT performs better than CDT in this scenario. If FDT performs better in general, then this differently expected utility calculation could be said to be a more accurate expected utility, which FDT is calculating in general.

On this first round, the average guess will be $\frac{1}{3} \cdot 50 + \frac{1}{3} \cdot 23.68 + \frac{1}{3} \cdot 21.05 = 31.58$. Two-thirds of this is 21.05, so FDT is precisely on target, and will perform the best.

Let us put our final evolutionary environment into practice. We will set our population at 10,000 agents, evenly split between each type, allowing a nice distribution of random numbers. We run the simulation for 100 rounds. Here is one run over 50 generations.

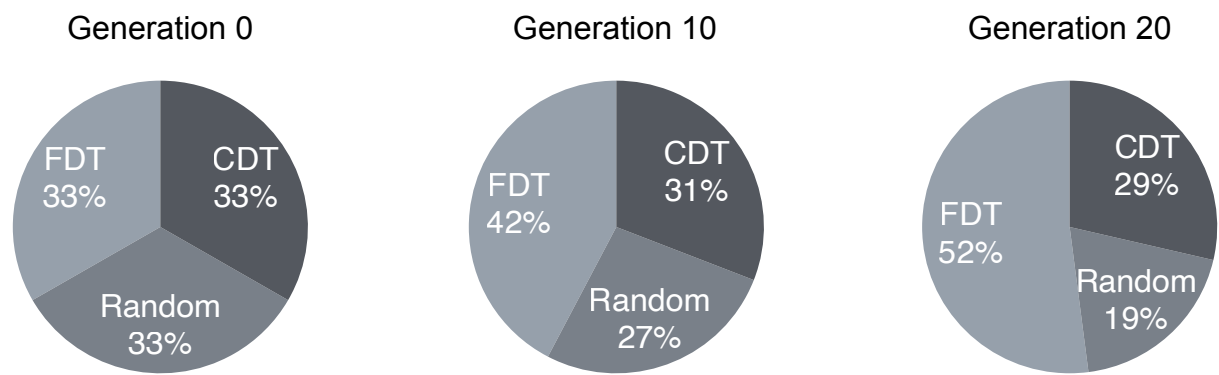

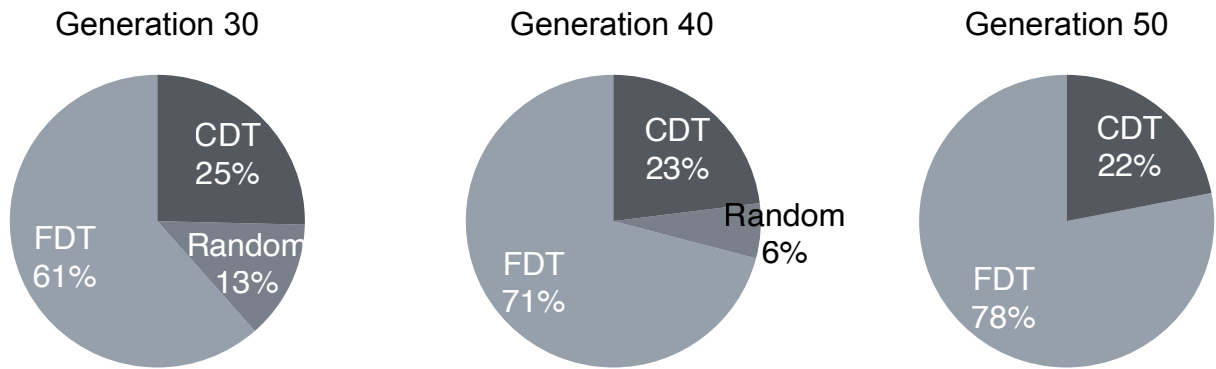

We have no payoffs to vary here, nor do we have a correlation value to vary. The only significant thing we can vary are the initial population rates. So let us take a look at that. We initialize CDT to be 80% of the population, while Random and FDT make up 10% each. We run this setup for 750 generations. Otherwise the parameters are the same.

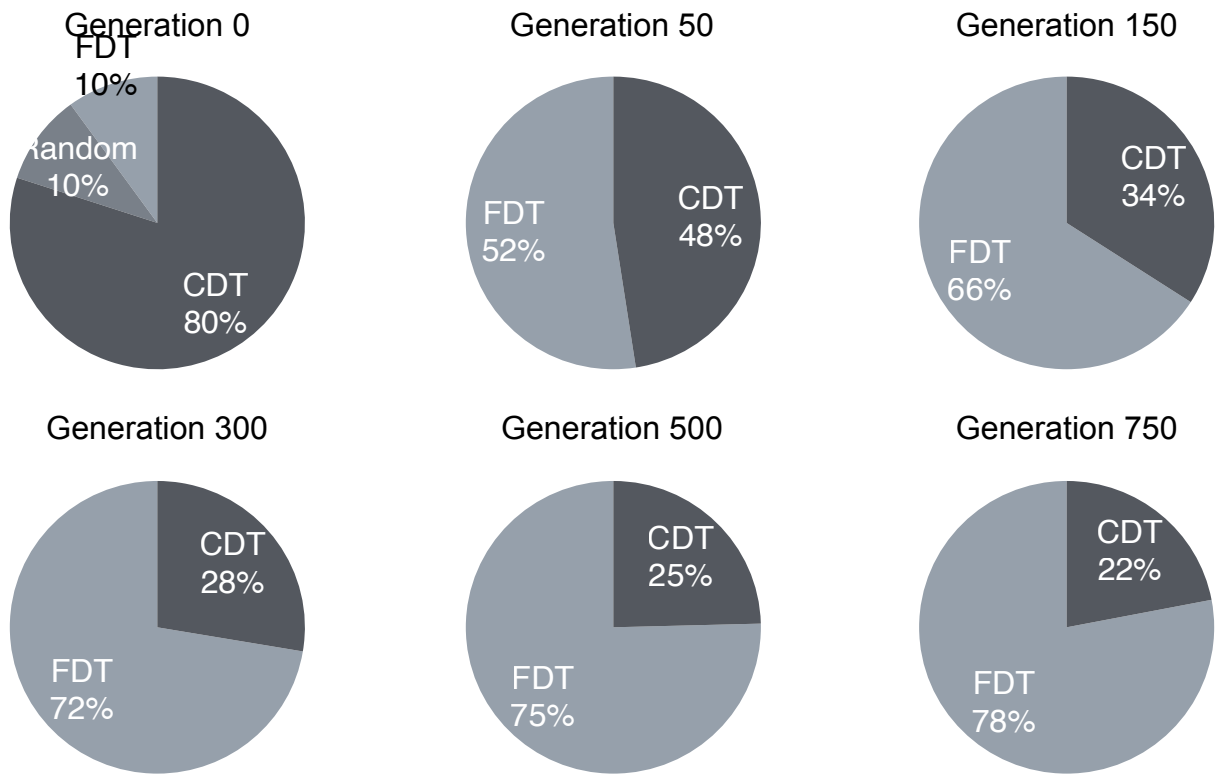

FDT appears to converge to exactly the same stopping point as before. There is little change after this point on both simulations. 78%/22% seems to be the natural stopping point for FDT and CDT in this game.

Part of this is because, when all the Random agents are gone, CDT attempts to guess 2/3 of the average of just the FDT agents, which is then the same as what FDT is doing, so they all guess 0, and FDT loses its advantage.

On the other hand, small mutations allow FDT to push higher when Random agents do crop up. Further, when FDT is very high in the population, they may be the main victims of mutation. These factors all seem to balance out at the given ratio. Only by removing all Random agents and mutation do CDT and FDT become evenly matched.

# 4 - Conclusions

## 4.1 Improving Upon This Research

Some improvements upon this research could certainly be made. A more refined construction of the Prisoner's Dilemma would have FDT agents perform the same action when they are in the same epistemic state, rather than when they receive the same signal. For instance, in our set up, we see the peculiar behavior that even when the population is 100% FDT, agents will still cooperate when they receive the signal that their opponent is FDT, and defect otherwise, even though in both cases they are still 100% certain that their opponent is an FDT agent. Really we would like agents who arrived at the same probability estimate in different ways to still output the same decision, but this is beyond the scope of our simple construction, which still demonstrates the main point.

In the Keynesian Beauty Contest, the behavior of CDT agents is perhaps rather shallow, as it is not really allowed to turn its causal modeling onto other CDT agents. Just because CDT agents assume that their behavior is independent of other CDT agents does not mean they cannot still model each other. It does seem, though, that however such modeling would work, it should include aggregating behavior from many agents, as we do. The best CDT agents could do here would be to assume that their behaviors are *not* independent, since they are not. Short of this, we expect CDT's model to be imperfect, so our heuristic is simply an approximation technique.

This approximation in fact helps CDT on at least some part of the game. In the final stage, when all Random agents have been eliminated, both CDT and FDT guess zero thereafter.

This is because CDT agents now make their predictions based only on FDT agents. Before this point, FDT is outperforming CDT, but at this point, they perform equally. If CDT agents actually followed a specific causal model of other CDT agents, we would expect this model to be imperfect, and FDT agents would continue to outperform them forever. Still, perhaps some more accurate heuristic could be used.

## 4.2 Further Research

The main longstanding open area of research in functional decision theory is how to construct logical diagrams from data in the first place. While computer scientist Judea Pearl has developed methods to construct causal diagrams from data, we currently have no known process to perform the analogous task for detecting logical connections among data. Can we see any path forward for this?

Here is a possibly related question: why do neural networks outperform causal networks? In machine learning, neural networks are trained on data to automatically establish weights and connections amongst input data to produce desired outputs. Causal networks are meant to perform a similar task, taking in data, feeding it through the weights and connections of the network, and outputting a desired action. However, causal networks are usually more hand-made. If causal connections are so important to understanding the world, why do neural networks perform so much better?

Obviously neural networks have the advantage of automatically finding and exploiting connections that human programmers are not aware of. But we might expect some of these con-nections to be rather spurious, more correlation than causation. One might call them evidential connections. For example, a neural network trained to identify pictures of dumbbells ended up thinking dumbbells had arms attached to them (Mordvintsev et al. 2015).

Neural networks end up forming a graph with vast numbers of variables connected in patterns that certainly do not form a true causal model of the world, and yet may be highly effective anyway.

If these connections could be pared down to strictly causal connections, we might expect even higher performance. Similarly, if we could make a neural network that automatically finds causal connections, the process may be able to find logical connections as well. That is, the neural network may be able to identify that some variables are dependent on a common decision function. The distillation of neural networks in ways like this is an active area of research in interpretable AI.

## 4.3 Philosophical Implications

How is this work relevant to the real world? Obviously there is much discussion of AI here. We are seeking the ideal theory of rationality, which would likely be useful in the construction of a machine that we wish to make intelligent decisions. If we do end up in a world with general AI, many of the issues of predictability and correlation of decision algorithms become especially relevant.

If an AI's source code can be accessed by a human or another AI, then their decision process could truly be highly transparent, and we have reason to believe that an AI using only classi-cal, causal reasoning would not perform well in this environment. The same holds for the possibility of AIs that really are copies of each other acting in the real world, or for AIs running similar software: they should be aware of their logical connections.

But is this research only relevant to a potential science-fiction future, or does it also have relevance for human decision making? We have tried to give plausible examples that imply that if humans are modestly transparent in their decisions, and humans' decisions are modestly correlated, then functional decision theory becomes relevant. We attempted to show that even if these correlations are not decision-relevant, they are frequently utility-relevant. One could then plausibly argue that this reasoning is relevant to nearly *all* human decisions.

One interesting implication of functional decision theory is that it seems worth spreading. The more people know it, the more we can trust that people are reasoning alike, and the better we can all perform. The same does not appear to be true of spreading traditional, causal decision theory. Many claim that studying economics and game theory makes people behave more selfishly (Gerlack 2017).

It seems to be harder to cooperate and behave pro-socially when one knows about game theory, and this can leave everyone worse off. But the more people that follow functional decision theory, the better off everyone appears to be. One would want their friends to know about the ideal theory of rationality, so that one could trust and rely on them.

As a final case study of a specific human-relevant problem, take voting. Causally speaking, the odds of your vote in an election making a difference is quite low. When a CDT agent imagines voting or not voting, it imagines all other votes remaining stable. Thus the utility of voting may seem quite low, and may not be worth the effort it takes to do so. Causal decision theory therefore rates virtually all votes as essentially pointless. But how can this be true, when we know that all the votes aggregate to something significant?

What does common sense say about voting? People commonly counter the argument above by saying "If everyone thought like that, then nobody would vote!" Functional decision theory aligns with this counterargument. When an FDT agent imagines voting or not voting, it counterfactually imagines that other voters shift along with it. The more similar these other voters are to the agent, the more likely they are to shift as well. By choosing to not vote, the agent has made it so that the agents most similar to itself are the least likely to vote, logically speaking.

This seems very counterproductive. If one wants their ideas to be represented, then ensuring that others with the most similar ideas do not vote is disastrous! Functional decision theory can therefore give a straightforward, rational reason to vote, assuming one wants others who share their ideas to vote as well.

We hope all this gives an idea of the applicability and importance of FDT. We believe it to be superior to CDT in real, human problems, as well as future problems related to artificial intelligence. We also believe it is closer to the ideal theory of rationality than CDT is, and is therefore a better model from which to build or think about an AI. How to do this in practice is, of course, yet to be determined.